\documentclass[12pt]{article}
\usepackage{mathrsfs}
\usepackage{graphicx,setspace,lscape,longtable}
\usepackage{amsmath,amsthm,amssymb,color}
\usepackage{epsfig}
\usepackage{natbib}
\usepackage{mathtools}
\usepackage{algpseudocode}
\usepackage{algorithm}
\usepackage{stmaryrd}
\usepackage{multirow}
\usepackage{url}
\usepackage{xurl}
\usepackage{xr}
\graphicspath{{./plot/}}
\usepackage{rotating}

\newcommand{\blind}{0}

\newtheorem{theorem}{Theorem}

\theoremstyle{remark}
\newtheorem{remark}{Remark}

\def\argmin{\mbox{argmin}}
\def\argmax{\mbox{argmax}}

\def\bfy{\mathbf{y}}

\def\bttau{\boldsymbol{\tilde{\tau}}}
\def\bhtau{\boldsymbol{\hat{\tau}}}
\def\btau{\boldsymbol{\tau}}
\def\epBIC{\text{ep-BIC}}

\begin{document}

\def\spacingset#1{\renewcommand{\baselinestretch}%
{#1}\small\normalsize} \spacingset{1}

\if0\blind
{
  \title{\bf Graph-based multiple change-point detection}
  \author{Yuxuan Zhang and Hao Chen\thanks{
    The authors were supported in part by the NSF Grant DMS-1848579.}\hspace{.2cm}\\
    \ \\
    Department of Statistics, University of California Davis, Davis, CA\\
   }
   \date{}
  \maketitle
} \fi

\if1\blind
{
  \bigskip
  \bigskip
  \bigskip
  \begin{center}
    {\LARGE\bf Graph-based multiple change-point detection}
\end{center}
  \medskip
} \fi

\bigskip
\begin{abstract}
We propose a new multiple change-point detection framework for multivariate and non-Euclidean data. First, we combine graph-based statistics with wild binary segmentation or seeded binary segmentation to search for a pool of candidate change-points. 
We then prune the candidate change-points through a novel goodness-of-fit statistic.
Numerical studies show that this new framework outperforms existing methods under a wide range of settings. The resulting change-points can further be arranged hierarchically based on the goodness-of-fit statistic. The new framework is illustrated on a Neuropixels recording of an awake mouse.
\end{abstract}

\noindent%
\vfill

\newpage
\spacingset{1.5}
\section{Introduction}
\label{sec:intro}

Change-point analysis is a long-established statistical topic but encounters enormous challenges in this century. In the era of big data, data are usually of high dimension and complexity. For example, in bioinformatics, finding DNA copy number variants in dozens to hundreds of samples is of scientific interest \citep{zhang2010, jiang2015}. In astrophysics, experts discover the presence of the galaxies using hyper-spectral data cubes obtained by integral field spectrograph \citep{enikeeva2019}. There is also a need in finding abrupt changes in dynamic communication networks, such as to investigate the relationship between communication patterns and external shock over time \citep{peel2015,dong2020,braun2021}.

Consider a sequence of independent observations $\left\{\mathbf{y}_{i}: i=1,\dots,n\right\}$, indexed by time or some other meaningful orderings, such that
\begin{equation}
\label{model}
\mathbf{y}_{i} \sim F_{j}, \quad \tau_{j}+1 \leq i \leq \tau_{j+1}, \quad j=0, \ldots, m,
\end{equation}
where $0=\tau_{0}<\tau_{1}<\cdots<\tau_{m+1}=n$, and $F_j$'s are arbitrary unknown probability measures, satisfying $F_j\neq F_{j+1}$. The parameters $\boldsymbol{\tau}=\left(\tau_{1}, \tau_{2}, \ldots, \tau_{m}\right)$ are the change-points of the process. Our goal is to estimate $m$ and $\boldsymbol{\tau}$.

Most existing works for multiple change-point detection with multivariate observations are based on parametric models. For example, \citet{zhang2010} and \citet{enikeeva2019} considered $\ell_{2}$ aggregation of cumulative sum statistic (CUSUM). \citet{cho2015} developed a truncated CUSUM combined with binary segmentation to tackle the sparsity in high dimensional data. \citet{wang2016} studied a projected CUSUM procedure also under a sparse high dimensional setting. \cite{lavielle2006} introduced a set of methods based on penalized Gaussian log-likelihood to detect changes in covariance structure. \citet{wang2018} improved Pearson's chi-squared test for multinomial data, and added a penalty term to allow for multiple change-point selection.

In recent years, more nonparametric methods are developed to avoid model misspecification in parametric methods. For example, \citet{matteson2014} proposed E-Divisive that combined Euclidean-based divergence measure and divisive algorithm. \citet{harchaoui2009} and \citet{arlot2019} used kernel-based statistics to measure the discrepancy between segments (KCpA and KCP). Another framework for multivariate and non-Euclidean data is the graph-based method proposed by \cite{chen2015}.
For the first time, it gives an \emph{analytic $p$-value approximation} for a nonparametric framework that can be applied to data in an arbitrary dimension or non-Euclidean data, facilitating its application to large data sets.
\citet{chu2019} improved the graph-based method by introducing new graph-based statistics that perform well under a wide range of alternatives.
However, unlike E-Divisive and KCP, the existing graph-based methods focused on the single change-point alternative and the changed interval alternative.

In this paper, we work out a reliable way of finding multiple change-points utilizing the graph-based statistic (Section \ref{framework}). In particular, we first adopt the idea of wild binary segmentation (WBS) \citep{fryzlewicz2014} and seeded binary segmentation (SBS) \citep{kovacs2020} to find a pool of candidate change-points. We then propose a pseudo-BIC criterion for change-point selection. Simulation studies show that this new framework has superb performance compared to other state-of-the-art methods under a variety of settings (Section \ref{simulation}). The new approach is illustrated by analyzing a Neuropixel dataset where multiple types of changes are found (Section \ref{realdata}).

\section{gMulti: A Graph-based Multiple Change-point Detection Framework}
\label{framework}
The proposed graph-based multiple change-point detection framework consists of two steps: searching and pruning. We begin by introducing some notations.

\subsection{Notations and Graph-based Single Change-point  Detection}
Here, we consider the scenario of detecting a single change point on $\{\bfy_{i}:a\leq i \leq b\}$, i.e., testing the null hypothesis $H_{0}^{[a,b]}: \mathbf{y}_{i} \sim F_{0}, \  i=a, \dots, b$ against the alternative $H_{1}^{[a,b]}: \exists \ a \leq \tau<b, \ \mathbf{y}_{i} \sim F_{0} \ $ for $a\leq i \leq \tau $ and $ \mathbf{y}_{i} \sim F_{1}\   \text {otherwise.}$
When the null hypothesis $H_0^{[a,b]}$ is true, we could arbitrarily permute the order of observations while keeping the joint distribution the same. So the permutation null distribution that places $1/(b-a+1)!$ probability on each of the $(b-a+1)!$ permutations of  $\{\bfy_{i}:a\leq i \leq b\}$ is used as a surrogate for the true null distribution. 

Let $G^{[a,b]}$ be the similarity graph on $\{\bfy_{i}:a\leq i \leq b\}$. The graph $G^{[a,b]}$ is an unweighted undirected acyclic graph within which edges are constructed based on a distance measure defined on the sample space up to a criterion. 
Some examples of $G^{[a,b]}$ include \textit{k}-minimum spanning tree (\textit{k}-MST) and \textit{k}-nearest neighbor graph (\textit{k}-NNG) (see \citet{chen2013, chen2015} for more discussions on choices of the graph). We use $G^{[a,b]}$ to denote both the graph and the set of edges in the graph when its vertex set is implicitly obvious.
Let $R^{[a,b]}_1(t)$ be the number of edges connecting observations within $[a,t]$, and $R^{[a,b]}_2(t)$ be the number of edges that connect observations within $[t+1, b]$.
The generalized edge-count scan statistic proposed in \cite{chu2019} is defined as:
\begin{equation}
\label{Sscan}
	\max _{n_{le}^{[a,b]} \leq t \leq n_{ri}^{[a,b]}} S^{[a,b]}(t),
\end{equation}
where
\begin{equation}
	S^{[a,b]}(t)=\left(\begin{array}{l}R^{[a,b]}_{1}(t)-\mathbf{E}\left(R^{[a,b]}_{1}(t)\right) \\ R^{[a,b]}_{2}(t)-\mathbf{E}\left(R^{[a,b]}_{2}(t)\right)\end{array}\right)^{T} \left(\boldsymbol{\Sigma}^{[a,b]}(t)\right)^{-1}\left(\begin{array}{l}R^{[a,b]}_{1}(t)-\mathbf{E}\left(R^{[a,b]}_{1}(t)\right) \\ R^{[a,b]}_{2}(t)-\mathbf{E}\left(R^{[a,b]}_{2}(t)\right)\end{array}\right),
\end{equation}
with $\Sigma^{[a,b]} (t)=\mathbf{Var}\left((R^{[a,b]}_{1}(t), R^{[a,b]}_{2}(t))^T\right)$.  The expectation and variance are defined under the permutation null distribution. Here, $n_{le}^{[a,b]}$ and $n_{ri}^{[a,b]}$ are pre-specified endpoints for the scan. 
In the following, we use $\lceil a+0.1(b-a+1)\rceil$ and $\lfloor b-0.1(b-a+1)\rfloor$ as default choices for $n_{le}^{[a,b]}$ and $n_{ri}^{[a,b]}$, respectively, where $\lceil \, \cdot \, \rceil$ is the ceiling function, and $\lfloor \, \cdot \,\rfloor$ is the floor function. We will focus on the generalized edge-count statistic in this paper as it considers a useful pattern for high-dimensional data and works well for a wide range of alternatives \citep{chen2017}. \citet{chu2019} also provided an analytic $p$-value approximation for the test statistics (\ref{Sscan}), and we  denote it by $\hat p(\{\bfy_{i}:a\leq i \leq b\})$ in the following.

\subsection{Step 1: Candidate Change-point Search}
\label{search}
We adapt the idea of WBS and SBS to construct the pool of candidate change-points. The pseudocodes are provided in Algorithm \ref{alg:gWBS} (\textsc{g.WBS}) and \ref{alg:gSBS} (\textsc{g.SBS}), respectively. Here are some notations used in the algorithms. Let $\alpha$ be the pre-specified significance level and MinLen be the minimum length of generated intervals. The number of randomly generated intervals in \textsc{g.WBS} is $L$, and the decay parameter used for \textsc{g.SBS} is $\gamma$.

\begin{algorithm}
\caption{Change-points search by graph-based WBS}
\label{alg:gWBS}
\begin{algorithmic}
		\Procedure{g.WBS}{$a$, $b$, $\bttau$, $\alpha$, $L$, MinLen} 
		\If{$b-a+1<\text{MinLen}$} 
		
		STOP
		\EndIf
		\If{$L\geq(b-a-\text{MinLen}+2)(b-a-\text{MinLen}+3)/2$}
		\State $L \gets (b-a-\text{MinLen}+2)(b-a-\text{MinLen}+3)/2$
		\State Draw all intervals $\left[a_{l}, b_{l}\right] \subseteq[a,\ldots, b]$, $l=1,\ldots , L$, s.t. $b_l-a_l+1 \geq \text{MinLen}$ 
		\Else
		\State Randomly and uniformly draw intervals $\left[a_{l}, b_{l}\right] \subseteq[a,\ldots, b]$, $l=1,\ldots , L$, s.t. $b_l-a_l+1 \geq \text{MinLen}$.
		\State Add $\left[a_0, b_0\right]=\left[a,b\right]$ to the set of intervals.
		\EndIf
		\State $l' \gets \argmin_{l \in \{0,\dots,l \}} \  \hat p(\{\bfy_{i}:a_l\leq i \leq b_l\})$
		\State $\hat{t}\gets \argmax_{n_{le}^{[a_{l'}, b_{l'} ]} \leq t \leq n_{ri}^{[a_{l'}, b_{l'} ]}} \ S^{[a_{l'}, b_{l'}]}(t)$
		\If{$ \hat p(\{\bfy_{i}:a_{l'}\leq i \leq b_{l'}\})< \alpha$}
		\State Add $\hat t$ to the set $\bttau$.

		\State \Call{g.WBS}{$a$, $\hat t$, $\bttau$, $\alpha$, $L$, MinLen}
		\State \Call{g.WBS}{$\hat t +1$, $b$, $\bttau$, $\alpha$, $L$, MinLen}
		\Else
		
		STOP
		\EndIf
		\EndProcedure
	\end{algorithmic}
	\end{algorithm}
The function \textsc{g.WBS} or \textsc{g.SBS} can be applied recursively to find the candidate change-points.
The function \textsc{g.WBS} starts by applying the generalized edge-count statistic to $L$ randomly generated intervals. If the smallest $p$-value among $L$ of them is less than the significance level $\alpha$, we add the corresponding detected change-point into $\bttau$ and continue by applying the function to the subsegments.
All potential intervals will be scanned if the subsequence $\left\{\mathbf{y}_{i}: a\leq i \leq b\right\}$ is too short to draw $L$ different intervals longer than $\text{MinLen}$. 
Rather than drawing all intervals initially like the traditional WBS, \textsc{g.WBS} draws a small number of new intervals as recursion proceeds. The traditional WBS requires a higher number of intervals to guarantee change-points are appropriately covered, but many generated intervals are too long and inefficient. 
The recursive drawing in \textsc{g.WBS} is similar to WBS2 proposed by \citet{fryzlewicz2020}. However, WBS2 will continue until all indices are included as potential change-points.

The function \textsc{g.SBS} uses a deterministic scheme to generate intervals so that it is more efficient than the traditional WBS. Following the recommendation in \citet{kovacs2020}, the collection of seeded intervals used in $\textsc{g.SBS}$ is
\begin{equation*}
	\mathcal{I}_\gamma=\bigcup_{k=1}^{\lfloor \log_{\gamma} \frac{\text{MinLen}-1}{n} +1 \rfloor} \bigcup_{j=1}^{2\lceil (1/\gamma)^{k-1}\rceil-1} \{ \left[ \lfloor(j-1)s_k \rfloor, \lceil(j-1)s_k+n\gamma^{k-1} \rceil \right]\},
\end{equation*}
where $s_k=n(1-\gamma^{k-1})/(2\lceil 1/\gamma^{k-1}\rceil-2)$.

\begin{algorithm}
\caption{Change-points search by graph-based SBS}\label{alg:gSBS}
\begin{algorithmic}
		\Procedure{g.SBS}{$a$, $b$, $\bttau$, $\alpha$, $\mathcal{I}_\gamma$} 
		\If{$b-a+1<\text{MinLen}$} 
		
		STOP
		\EndIf
		\State $\mathcal{M}_{a,b} \gets$ set of indices $l \in \mathcal{I}_\gamma$ such that $\left[a_{l}, b_{l}\right] \subseteq [a, \ldots, b]$
		\State $\mathcal{M}_{a,b}\gets\mathcal{M}_{a,b} \cup\{0\}$, where $\left[a_0, b_0\right]=\left[a,b\right]$
		\State $l' \gets \argmin_{l\in \mathcal{M}_{a,b}} \  \hat p(\{\bfy_{i}:a_l\leq i \leq b_l\})$
		\State $\hat{t}\gets \argmax_{n_{le}^{[a_{l'}, b_{l'} ]} \leq t \leq n_{ri}^{[a_{l'}, b_{l'} ]}} \ S^{[a_{l'}, b_{l'}]}(t)$
		\If{$ \hat p(\{\bfy_{i}:a_{l'}\leq i \leq b_{l'}\})< \alpha$}
		\State Add $\hat t$ to the set $\bttau$.
		\State \Call{g.SBS}{$a$, $\hat t$, $\bttau$, $\alpha$, $\mathcal{I}_\gamma$}
		\State \Call{g.SBS}{$\hat t +1$, $b$,  $\bttau$, $\alpha$, $\mathcal{I}_\gamma$}
		\Else
		
		STOP
		\EndIf
		\EndProcedure
	\end{algorithmic}
	\end{algorithm}

In this step, we aim to find all potential change-points, so we would like to be more inclusive. Theoretically speaking, larger values of $\alpha$, $\gamma$, and $L$ would bring in more candidate change-points, but also result in a longer computation time. Investigators can set those parameters according to their needs.
The default value of $L$ is set to be 100, and the default value for $\alpha$ is 0.01. This is in general enough for $n$ up to a few thousand. If $n$ is even larger, then $L$ can also be set larger.
For \textsc{g.SBS}, we adopt the recommendation in \cite{kovacs2020} and set  $\gamma=\sqrt{0.5}$. The value of MinLen affects the power of detecting frequent changes. The functions \textsc{g.WBS} and \textsc{g.SBS} could detect change-points that are at least $\text{MinLen}/2$ apart from each other. We set $\text{MinLen}$ to be 10 as the default choice, which is usually enough even for cases with frequent changes.

We have no intention to compare the two methods in detail since they are both reliable in general. One may choose between them according to their needs and understandings. From our experience, \textsc{g.SBS} has comparable power and faster speed compared with \textsc{g.WBS} when the data is not complex. However, \textsc{g.WBS} shows better power when the change-points are frequent, as it scans on more intervals when the subsequence $\left\{\mathbf{y}_{i}: a\leq i \leq b\right\}$ is short.

\subsection{Step 2: Candidates Pruning}
\label{selection}
\subsubsection{Penalized adjacent sum statistic}
\label{passection}
In the second step, we introduce a goodness-of-fit statistic to perform candidate change-point pruning. Let $\bttau=\{\tilde \tau_1, \tilde \tau_2,\dots, \tilde \tau_{\tilde m}\}$ denotes the set of cadidate change-points found in step 1, where $1\leq\tilde\tau_1 <\tilde\tau_2<\dots<\tilde\tau_{\tilde m}\leq n-1$, and $\tilde m=\lvert\bttau \rvert$, where $\lvert \,\cdot\, \rvert$ is the cardinality of a set. We define $\bttau$'s corresponding boundary set $\boldsymbol{\tilde{\eta}}$ as
\begin{equation*}
	\boldsymbol{\tilde{\eta}}=\{\tilde\eta_0,\tilde\eta_1,\dots, \tilde\eta_{\tilde m}\}\coloneqq
	\begin{cases}
		\{0,n\}& \text{if } \tilde m=1,\\
		\{0,\lceil \frac{\tilde\tau_1+\tilde\tau_2}{2}\rceil,\lceil \frac{\tilde\tau_2+\tilde\tau_3}{2}\rceil,\dots,\lceil \frac{\tilde\tau_{\tilde m-1}+\tilde\tau_{\tilde m}}{2}\rceil  ,n\}&\text{if } \tilde m\geq2.
	\end{cases}
\end{equation*}
We define an adjacent sum goodness-of-fit statistic
\begin{equation}
	AS(\bttau)=\sum_{j=1}^{\tilde m}S^{\left[\tilde\eta_{j-1}+1, \tilde\eta_{j} \right]}(\tilde \tau_j).
\end{equation}
Each summand in $AS(\bttau)$ is  a local two-sample test statistic measuring credibility of a candidate change-point $\tilde \tau_j$. The subsample used in each $S^{\left[\tilde\eta_{j-1}+1, \tilde\eta_{j} \right]}(\tilde \tau_j)$ starts from $\tilde \eta_{j-1}+1$, i.e., the middle point of $\tilde \tau_{j}$ and $\tilde \tau_{j-1}$. It then ends at $\tilde \eta_{j}$, i.e., the middle point of $\tilde \tau_{j}$ and $\tilde \tau_{j+1}$. If $\tilde \tau_j$ is a true change-point, it will lead to a relatively large $S^{\left[\tilde\eta_{j-1}+1, \tilde\eta_{j} \right]}(\tilde \tau_j)$; and vice versa.
\begin{figure}[H]
	\centering
	\includegraphics{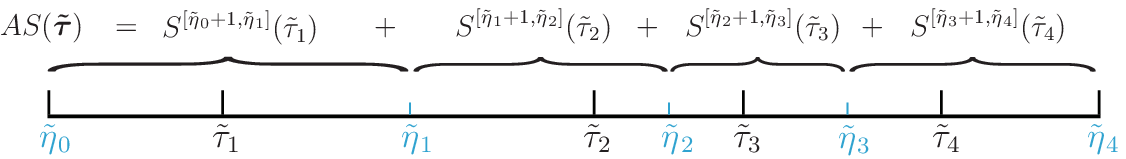}
	\caption{\footnotesize An illustration of how $AS(\bttau)$ is calculated with four candidate change-points.}
	\label{fig:step2}
\end{figure}

\begin{figure}[h]
	\centering
	\includegraphics[width=1\linewidth]{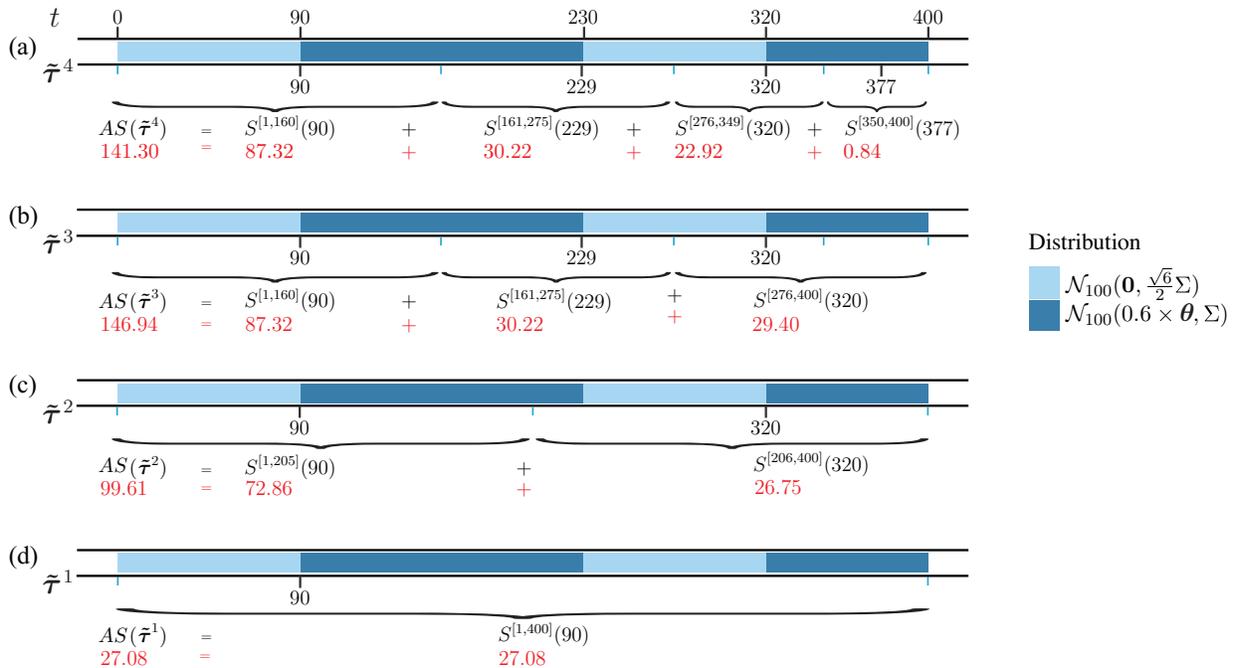}
	\caption{\footnotesize $AS(\bttau)$ on four possible change-points sets $\bttau$. The sequence is of 400 observations and 3 change-points at 90, 230, and 320. The first and third subsequences have the same distribution while the others have different mean and covariance structures. In (a), $\bttau^4$ overfits the data. In (b), $\bttau^3$ is close to the true $\btau$, and $AS(\bttau)$ is maximized. In (c) and (d), underestimated $\tilde m$ lead to small adjacent sum values.}
	\label{fig:egstep2}
\end{figure}

We illustrate how $AS(\bttau)$ works through an toy example: a normally distributed sequence with $n=400$ and $\btau=\{90, 230, 320\}$. Let $\bfy_i \overset{\text{i.i.d.}}{\sim} \mathcal{N}_{100}(\mathbf{0},\frac{\sqrt6}{2}\Sigma)$ if  $1\leq i \leq 90$ and $231\leq i \leq 320$; $\bfy_i \overset{\text{i.i.d.}}{\sim} \mathcal{N}_{100}(0.6\times\boldsymbol\theta,\Sigma)$ otherwise, where $\boldsymbol{\theta}$ is a vector with the first 20 elements all ones and the rest zeros, and $\Sigma_{jk}=0.3^{\lvert j-k \rvert}$ for $1\leq j,k \leq 100$. 
In one simulation run, the candidate change-points resulted from step 1 are $\bttau^4=\{90,229,320,377\}$, with a falsely detected change-point 377. Now, $AS(\bttau^4)=141.30$ and 4 local statistics are shown in Figure \ref{fig:egstep2} (a). Among them, $S^{[350, 400]}(377)$ is the smallest, which is expected as 377 is not a true change-point. 
Meanwhile, $S^{[350, 400]}(377)$ retains observations its neighbor statistic $S^{[276, 349]}(320)$ can use.
If we remove 377 from $\bttau^4$, we see $\bttau^3$ is almost equivalent to $\btau$. $AS(\bttau^3)$ increases to 146.94. Although $S^{[350, 400]}(377)$ is removed, $S^{[276, 400]}(320)$ is larger than $S^{[276, 349]}(320)$ by more than that of $S^{[350, 400]}(377)$ (Figure \ref{fig:egstep2} (b)). If we further drop a true change-point, for example, 229 from $\bttau^3$, $AS(\bttau^2)$ quickly falls. Here, $S^{[206, 400]}(320)$ contains three segments and becomes small (Figure \ref{fig:egstep2} (c)).

Roughly speaking, when $\bttau$ contains false discoveries those $S^{\left[\tilde\eta_{j-1}+1, \tilde\eta_{j} \right]}(\tilde \tau_j)$'s with falsely detected $\tilde{\tau}_j$'s are usually small, and the existence of these falsely detected $\tilde{\tau}_j$'s would affect the values of $S^{\left[\tilde\eta_{j-2}+1, \tilde\eta_{j-1} \right]}(\tilde \tau_{j-1})$ and $S^{\left[\tilde\eta_{j}+1, \tilde\eta_{j+1} \right]}(\tilde \tau_{j+1})$ as the observations in $[\tilde\eta_{j-1}+1, \tilde\eta_{j}]$ are reserved for $\tilde{\tau_j}$. When $\bttau$ misses true change-points $AS(\bttau)$ would lose those portions contributed by those left-out true change-points. Also, those left-out true change-points might affect the remaining $\tilde{\tau}_j$'s in $\bttau$ as some corresponding intervals could contain more than two segments.

\begin{figure}[h]
	\centering
	\includegraphics[width=1\linewidth]{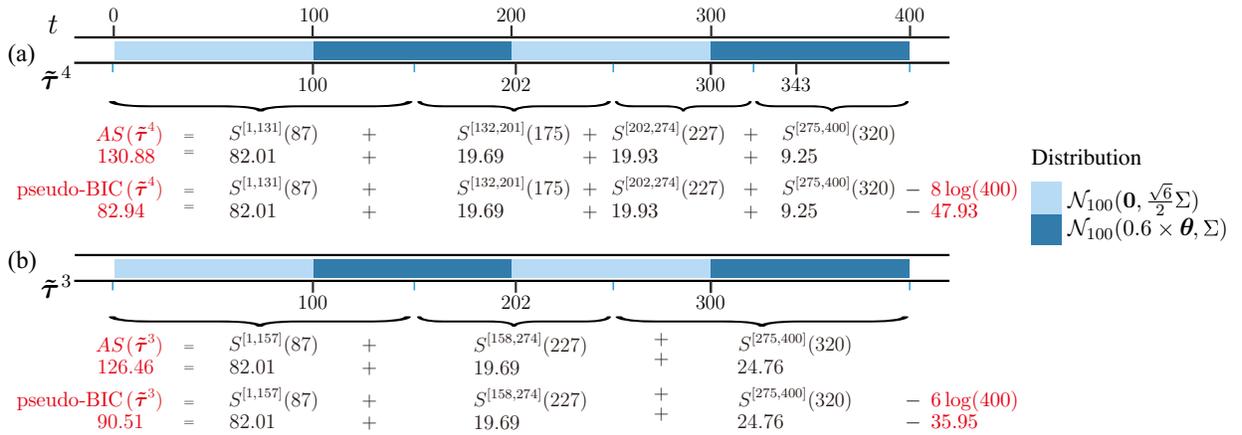}
	\caption{\footnotesize Comparison between $AS(\bttau)$ and $\text{pseudo-BIC}(\bttau)$ on two possible change-points sets $\bttau^4$ and $\bttau^3$. Here, $\text{pseudo-BIC}(\bttau^3)$ is greater than $\text{pseudo-BIC}(\bttau^4)$, while $AS(\bttau^3)$ is less than $AS(\bttau^4)$.}
	\label{fig:ASvsPseudoBIC}
\end{figure}

However, the mechanism of the adjacent sum may fail sometimes -- removing a false change-point may decrease adjacent sum as shown in Figure \ref{fig:ASvsPseudoBIC}. This is not surprising if we view it as a model selection problem and overfitting happens commonly. We thus introduce a penalty term to avoid overfitting. To determine the penalty term, we first study the null distribution of the adjacent sum.

Under the null that there is no change-point in the entire sequence, $AS(\bttau)$ has good asymptotic property. Let $G_i^{[a,b]}$ be a subgraph of $G^{[a,b]}$ containing all edges that connect to node $\bfy_i$, and $G_{i,2}^{[a,b]}$ be a subgraph of $G^{[a,b]}$ containing all edges with at least one endpoint in the node set of $G_i^{[a,b]}$. Then $\lvert G_i^{[a,b]}\rvert$ is the degree of node $\bfy_i$ in $G^{[a,b]}$. We also define $V_{G^{[a,b]}}$ as $V_{{G}^{[a,b]}}:=\sum_{i=a}^{b}\left|G_{i}^{[a,b]}\right|^{2}-\frac{4( |G^{[a,b]}|)^{2}}{b-a+1}$. The value  $V_{G^{[a,b]}}$ describes the variability of the sequence $|G^{[a,b]}_i|$'s. In the following, we write $u_n=O(v_n)$ when $u_n$ has the same order as $v_n$, and $u_n=o(v_n)$ when $u_n$ has order smaller than $v_n$.

\begin{theorem}
	\label{theorem1}
	For mutually disjoint intervals $[a_j,b_j]$ and $t_j \in [a_j,b_j]$, $j=1,\dots, m$, 
	when $\sum_{i=a_j}^{b_j}\lvert G_i^{[a_j,b_j]}\rvert^2=o(\lvert G^{[a_j,b_j]}\rvert^{\frac{3}{2}})$, 
	$\sum_{i=a_j}^{b_j}\lvert G_{i,2}^{[a_j,b_j]}\rvert^2=o(\lvert G^{[a_j,b_j]}\rvert V_{G^{[a_j,b_j]}})$, 
	$\sum_{i=a_j}^{b_j}\lvert G_i^{[a_j,b_j]}\rvert^3=o(V_{G^{[a_j,b_j]}}\sqrt{\min\{\lvert G^{[a_j,b_j]} \rvert, V_{G^{[a_j,b_j]}}\}})$, as $b_j-a_j\rightarrow\infty$, 
	and $t_j/(b_j-a_j)\rightarrow u_j$, $0<u_j<1$ holds for each $j$, we have
\begin{equation*}
		\sum_{j=1}^{m}S^{[a_j,b_j]}(t_j)\stackrel{d}{\rightarrow}\chi^2_{2m}.
\end{equation*}
\end{theorem}

\begin{remark}
Theorem \ref{theorem1} reveals that $AS(\bttau)$ is asymptotical $\chi^2_{2\tilde m}$ under the null hypothesis of no change-points for any fixed $\bttau$ with $\tilde m$ elements. Here the permutation null is within each interval $\left\{\bfy_i: \tilde\eta_{j-1}+1\leq i \leq \tilde\eta_{j} \right\}$ in $AS(\bttau)$ rather than the whole sequence $\left\{\bfy_i: 1\leq i \leq n \right\}$. Therefore, it is more rigorous to say $AS(\bttau)$ is asymptotical $\chi^2_{2\tilde m}$ under a group of permutation null $(H_0^{\left[1, \tilde\eta_1\right]}, H_0^{\left[\tilde\eta_1+1, \tilde\eta_2\right]},\dots, H_0^{\left[\tilde\eta_{\tilde m -1}+1, n\right]})$.
\end{remark}

Theorem \ref{theorem1} is a natural extension of Theorem 2.1 of  \citet{zhu2021}. This asymptotic distribution coincides with that of the likelihood ratio test under 1-dimensional Gaussian change-points scenario. To be specific, let $F_j$ in (\ref{model}) be a univariate Gaussian measure with an unknown mean $\mu_j$ and a known variance $1$. Let $\mathcal{M}_m$ be the Gaussian model with $m$ change-points. Correspondingly, $\mathcal{M}_0$ is the Gaussian model with no change-points in the whole sequence. It is easy to show that (see for example \cite{zhang2005dissertation}):
\begin{equation}
\label{lr}
	2\log\frac{\mathbf{P}(\bfy_1,\dots,\bfy_n \mid \mathcal{M}_m)}{\mathbf{P}(\bfy_1,\dots,\bfy_n \mid \mathcal{M}_0)} \sim \chi^2_{m}
\end{equation}
under the universal null $\mathcal{M}_0$. The corresponding BIC in selecting change-points is
\begin{equation}
	\label{BIC}
	2\log\frac{\mathbf{P}(\bfy_1,\dots,\bfy_n \mid \mathcal{M}_m)}{\mathbf{P}(\bfy_1,\dots,\bfy_n \mid \mathcal{M}_0)}-m \log n.
\end{equation}
The null distribution of (\ref{lr}) and the asymptotic distribution of $AS(\bttau)$ are both chi-square, but with different degrees of freedom.
Hence, we propose a pseudo-BIC for our framework:
\begin{equation}
\label{pBIC}
	\text{pseudo-BIC}(\bttau)=AS(\bttau)-2\tilde m\log n.
\end{equation}
The penalty term used in (\ref{pBIC}) is two times that in (\ref{BIC}), which is consistent with the degrees of freedom. With the added penalty term, $\text{pseudo-BIC}(\bttau)$ could better avoid overfitting. In Figure \ref{fig:ASvsPseudoBIC}, $AS(\bttau)$ overfits the data, while $\text{pseudo-BIC}(\bttau)$ relieves this problem. 

\begin{figure}[h]
	\centering
	\includegraphics[width=1\linewidth]{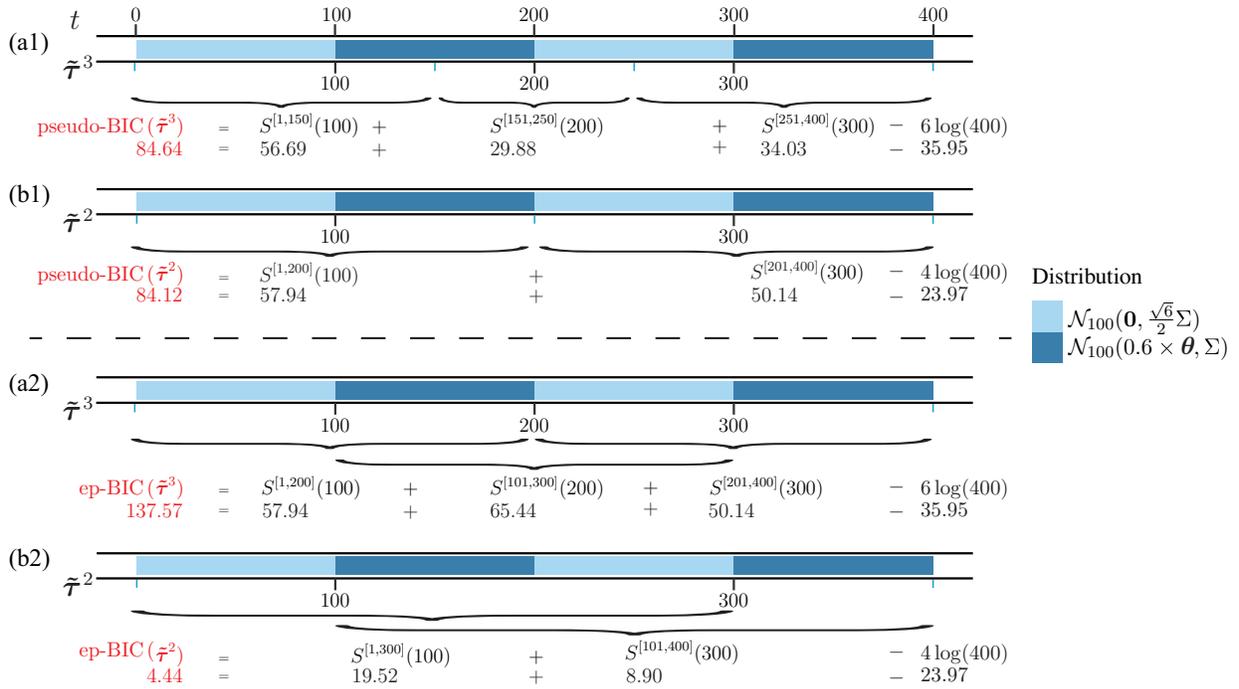}
	\caption{\footnotesize Comparison between $\text{pseudo-BIC}(\bttau)$ and $\text{ep-BIC}(\bttau)$. By using more information in each generalized edge-count statistic, ep-BIC is more likely to choose the correct model.}
	\label{fig:pseudoVSep}
\end{figure}

In adjacent sum and pseudo-BIC, each two-sample test statistic uses only half of the information of subsequence. This might cause power loss when the data is complex. Also, in some special cases, pseudo-BIC is not sensitive enough to cause underfitting. For example, in Figure \ref{fig:pseudoVSep}, $\text{pseudo-BIC}(\bttau^2)$ is almost equal to $\text{pseudo-BIC}(\bttau^3)$ because $S^{[1,200]}(100)$ and $S^{[201,400]}(300)$ happen to cover exactly two inhomogeneous subsequences. In Figure \ref{fig:pseudoVSep} (a1), all generalized edge-count statistics only cover part of the subsequences, causing power loss.
When pseudo-BIC is used in real applications, a more aggressive version is recommended. Define an expanded adjacent sum statistic as
\begin{equation}
	eAS(\bttau)=\sum_{j=1}^{\tilde m}S^{\left[\tilde\tau_{j-1}+1, \tilde\tau_{j+1} \right]}(\tilde \tau_j),
\end{equation}
where $\tilde\tau_0=0$ and $\tilde \tau_{\tilde m+1}=n$. This expanded version uses two times the information in each summand compared to the non-overlapped $AS(\bttau)$. When $\bttau$ is close to the true change-points $\btau$, $eAS(\bttau)$ will be greater than $AS(\bttau)$ (Figure \ref{fig:pseudoVSep} (a2)). On the other hand, when $\bttau$ misses some true change-points, $S^{\left[\tilde\tau_{j-1}+1, \tilde\tau_{j+1} \right]}(\tilde \tau_j)$ is more likely to cross true change-points and results in a small value. In this way, it better transforms the shortcoming of binary segmentation to an advantage here to avoid underfitting (Figure \ref{fig:pseudoVSep} (b2)). As for the corresponding pseudo-BIC, it is prudent to define the extended pseudo-BIC as
\begin{equation}
	\epBIC(\bttau)=eAS(\bttau)-c\tilde{m}\log n
\end{equation}
due to the introduced local dependence between $S^{\left[\tilde\tau_{j-1}+1, \tilde\tau_{j+1} \right]}(\tilde \tau_j)$. It is challenging to give an appropriate $c$ analytically due to the dependency.
Numerical results (Section \ref{penaltychoice}) show that $c=2$ seems to be reasonable under different settings, so we set the default choice of $c$ to be 2.
On average, each $S^{\left[\tilde\tau_{j-1}+1, \tilde\tau_{j+1} \right]}(\tilde \tau_j)$ leads to a penalty of $2\log n$ in (\ref{pBIC}). For that reason, $2\tilde{m}\log n$ is reasonable in the sense of first-order approximation. 

\subsubsection{Backward elimination}
\label{backward}

Given the set $\bttau$ of candidate change-points and the goodness-of-fit statistic $\epBIC(\bttau)$, the pruning of the change-points could be viewed as a model selection problem. We use backward elimination to fastly get the final set of pruned change-points $\bhtau$ as shown in Algorithm \ref{alg:gBE} (\textsc{g.BE}).
Starting from $\bttau^{\tilde m}$ estimated in step 1, \textsc{g.BE} evaluates how well different $\bttau^{l}$ fits the data. Finally, \textsc{g.BE} returns a sequence of $\epBIC(\bttau^l)$, and the $\bttau^l$ with the largest ep-BIC value is chosen as $\bhtau$. Note that \textsc{g.BE} stops until there is no change-point.
Together with the change-point dendrogram proposed in the following (Section \ref{hierarchicalstructure}), the sequence of $\epBIC(\bttau^l)$ provides investigators an ordered list of the change-points. 

\begin{remark}
	Given the nature of model selection, one may consider all subset approach, i.e., evaluating all possible subsets of $\bttau$ and choose the one with the best fit. This can be easily done when $\lvert\bttau\rvert$ is small. However, all subset approach can be computationally inhibitive in real applications where hundreds or thousands of change-points exist. 
\end{remark}

\begin{algorithm}
\caption{Backward elimination with $\epBIC$}
\label{alg:gBE}
\begin{algorithmic}
\Procedure{g.BE}{$\bttau$}
\State $l \gets \tilde m$
\State$\bttau^l\gets \bttau$
\While{$\lvert\bttau^l\rvert\geq1$}
\State $\mathbf{T}^l\coloneqq$ collection of change-points set $\bttau^l\backslash\{\tilde{\tau}^l_j\}$, where $\tilde{\tau}^l_j\in \bttau^l$, $j=1,\dots, l$
\State $\bttau^{l-1}\gets \argmax_{\boldsymbol{t}\in\mathbf{T}^l}\epBIC(\boldsymbol{t})$
\State $l\gets l-1$
\EndWhile
\State $\hat m\gets \argmax_l \epBIC(\bttau^{l})$
\State $\bhtau\gets \bttau^{\hat m}$
\State \textbf{return} $\bhtau$
\EndProcedure
\end{algorithmic}
\end{algorithm}

\subsection{Graph Choice}
\label{graphchoice}
Now we elaborate on the choices of similarity graphs used in the framework and their impact. From earlier works on graph-based tests \citep{friedman1979,chen2013,chen2017,chen2018}, $k$-MST is a recommended choice. 
However, the choice of $k$ is unsettled.
\citet{friedman1979,chen2017,chen2015,chu2019} showed that, for $k=O(1)$, larger $k$'s are preferred. \cite{zhu2021} further showed that $k$ of a higher order than $O(1)$ could even result in a higher power. We next discuss the choice of $k$ for the two steps separately.

In step 1, we want to be inclusive, and it is okay to have some false discoveries. So power is the main factor.
Without any prior knowledge about a new sequence, it is both intuitive and practical to choose $k$ depending on the length of the sequence. Notice that a $k$-MST built on $G^{[a_l,b_l]}$ contains $k(b_l-a_l)$ edges, while the information contained in the distance matrix is of order $O((b_l-a_l)^2)$. When $k=O(1)$, $k$-MST only uses a little information in the distance matrix. With this in mind, we consider $k=(b_l-a_l)^\lambda$, $0<\lambda<1$ to make use of more information. We compare the detection power for different $\lambda$'s under various simulation settings. To be specific, i.i.d. sequences with a single change-point $\tau=n/2$ were generated from three distribution pairs ($F_0$, $F_1$) in $\mathbb{R}^{d}$ (including mean shift and covariance structure change under multivariate normal and multivariate Cauchy distribution), each in three different dimensions ($d=20, 100, 500$) and four different lengths ($n=20, 50, 100, 200$). $\lfloor n^{
\lambda} \rfloor$-MST was constructed on the data based on the Euclidean distance. In each simulation, if $\hat p(\{\bfy_i:1\leq i \leq n\})<0.01$ and $\argmax_t S^{[1,n]}(t)\in [\tau-0.05n, \tau+0.05n]$, we deem it a successful detection. Detection power is defined as the proportion of successful detections.

\begin{figure}[p!]
	\centering
	\includegraphics[width=1\linewidth]{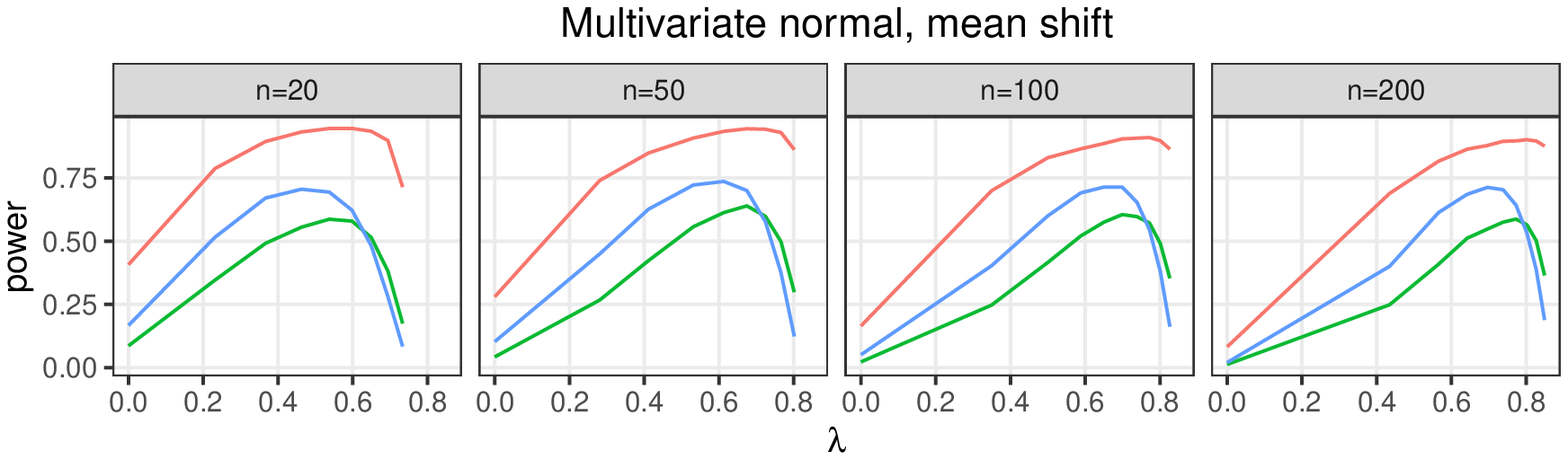}
	\includegraphics[width=1\linewidth]{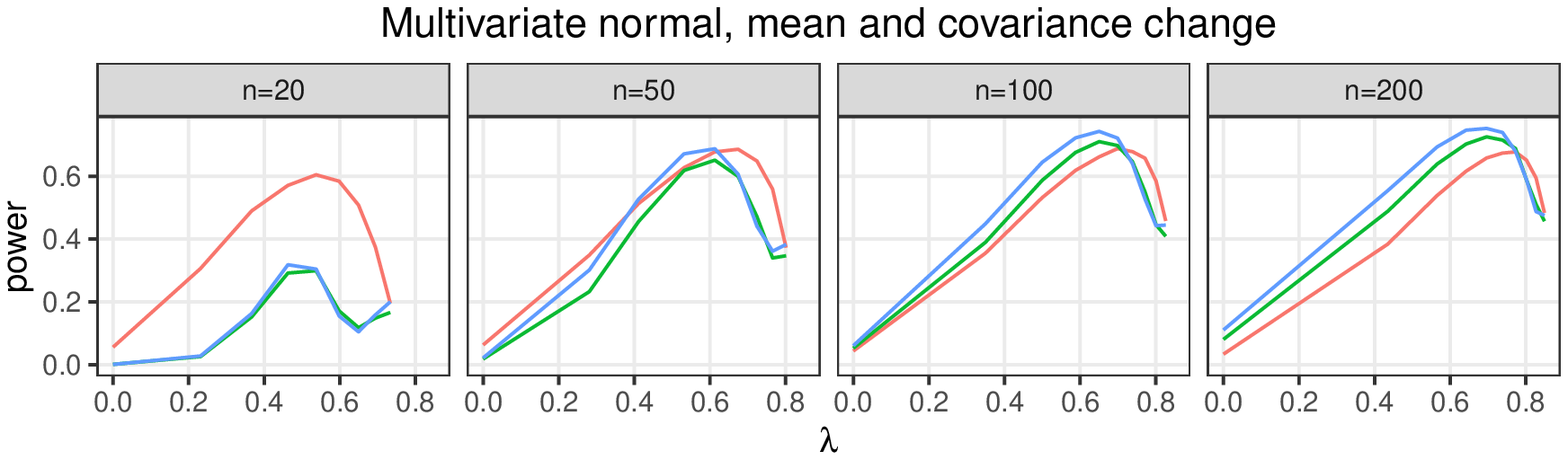}
	\includegraphics[width=1\linewidth]{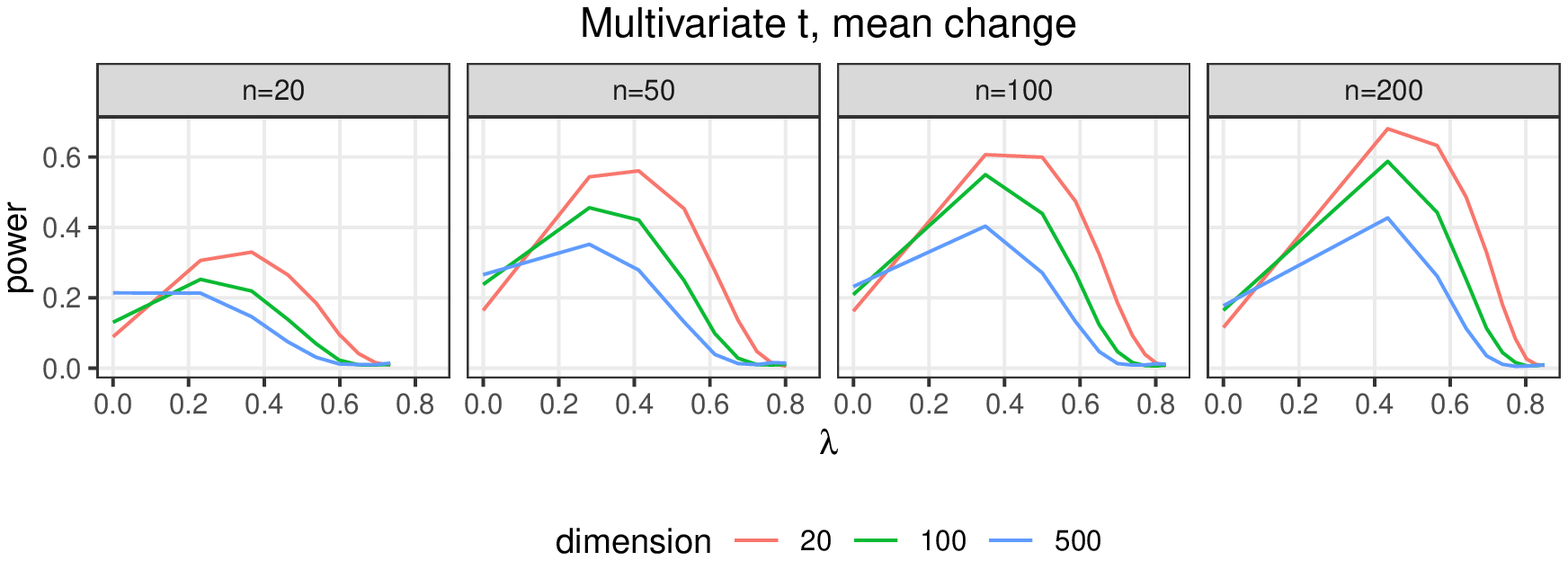}
	\caption{\footnotesize Detection power for singe change-point detection based on $\lfloor n^{\lambda} \rfloor$-MST. Simulation was replicated 5000 times under each setting. Three distribution pairs are ($\mathcal{N}_{d}(\mathbf{0},  \Sigma)$, $\mathcal{N}_{d}(\frac{20}{\sqrt{dn}}\mathbf{1}, \Sigma)$), ($\mathcal{N}_{d}(\mathbf{0},  \Sigma)$, $\mathcal{N}_{d}(\frac{15}{\sqrt{dn}}\mathbf{1}, (1+\frac{3}{2\sqrt n})\Sigma)$), and ($\text{Cauchy}_d(\mathbf{0},  I)$, $\text{Cauchy}_d(\frac{12}{\sqrt{n}\log d}\mathbf{1}, I)$), where $\Sigma_{j k}=0.3^{\mid j-k\mid}$ and $I$ is the identity matrix.}
	\label{fig:optiK}
\end{figure}

In the simulation, specific alternatives are chosen so that detection powers are comparable across different $\lambda$'s. The difference of detection power under different $n$'s and $d$'s does not reflect their influence on the method. From Figure \ref{fig:optiK}, we see that when $\lambda=0.5$, graph-based method shows adequate detection power across different simulation settings, though the optimal $\lambda$ varies between 0.3 to 0.7. According to this empirical result, we use $\min (30,\lfloor \sqrt{b_l-a_l} \rfloor)$-MST as the default similarity graph in Algorithm \ref{alg:gWBS} and \ref{alg:gSBS}. The upper bound 30 is set merely for computational consideration for very long sequences. 

For step 2, however, comparability of $\text{ep-BIC}(\bttau)$ on different $\bttau$'s is more essential. 
Usually, $\lvert \tilde\tau_{j+1}-\tilde\tau_{j-1}\rvert$ varies a lot for different $j$'s. Setting $k$ to be a function of $ \tilde\tau_{j+1}-\tilde\tau_{j-1}$ might lead to incomparable $S^{[\tilde{\tau}_{j-1}+1,\tilde{\tau}_{j+1}]}(\tilde \tau_j)$'s between long and short subsequences. Even when a signal is strong for a short subsequence, a small $k$ may not yield large $S^{[\tilde{\tau}_{j-1}+1,\tilde{\tau}_{j+1}]}(\tilde \tau_j)$. On the contrary, for a long sequence with a weak signal, a large $k$ may yield large statistic values, making ep-BIC prefer over-simplified models. To avoid this imbalance while keeping fine test performance, a constant $k-$MST, like the 5-MST, is preferred. Given that 5-MST can not be built on very short intervals, we set the default choice to be $\min(5,\lfloor\sqrt{ \tilde{\tau}_{j+1}-\tilde{\tau}_{j-1}}\rfloor)$-MST for step 2.

\subsection{Choice of $c$}
\label{penaltychoice}

\begin{figure}[h!]
	\centering
	\includegraphics[width=1\linewidth]{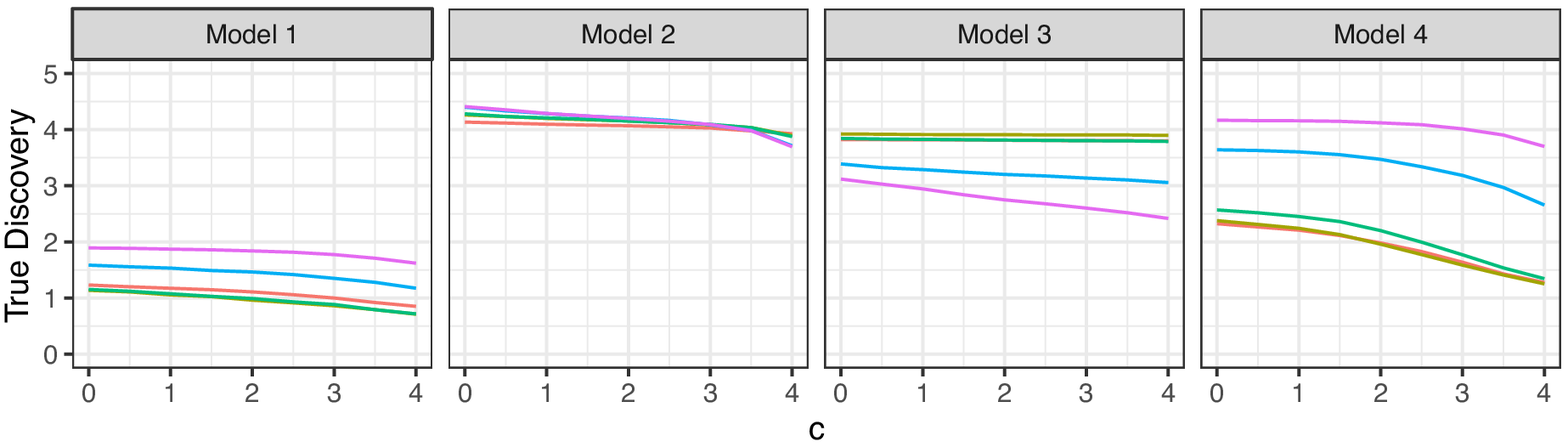}
		\includegraphics[width=1\linewidth]{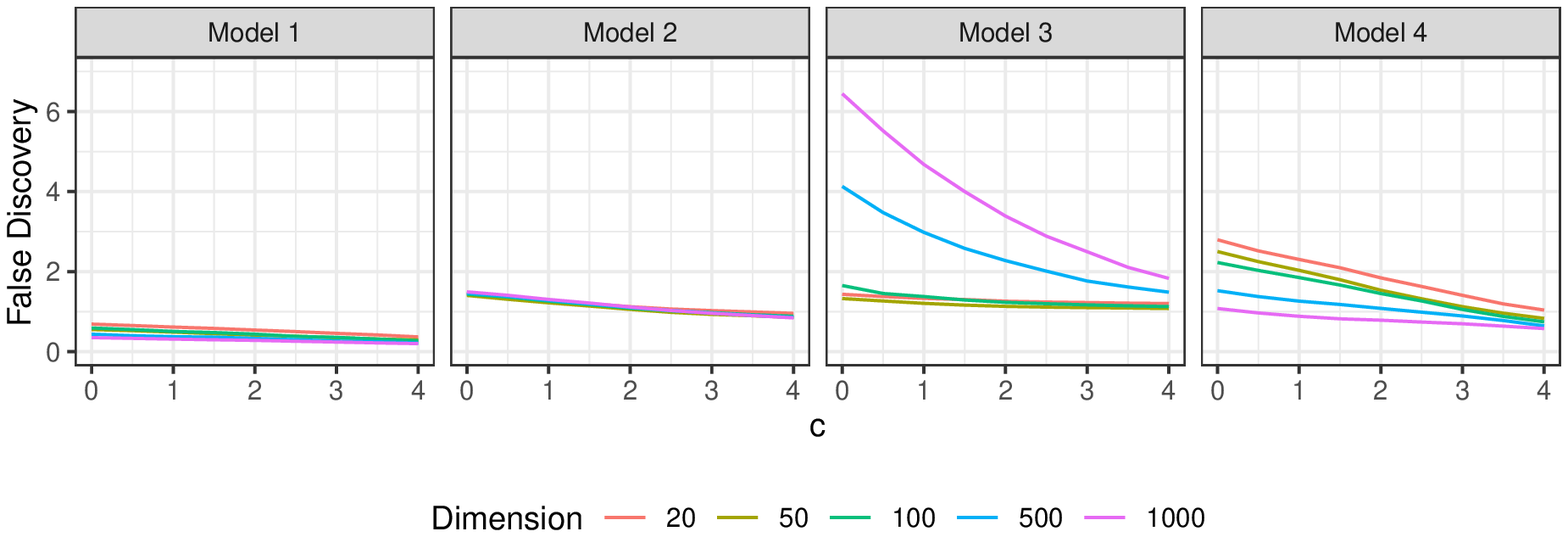}
	\caption{\footnotesize Average number of true discoveries and false discoveries under different penalties and settings with 1000 replications. }
	\label{fig:optiC}
\end{figure}

Recall that extended pseudo BIC is defined as:
\begin{equation*}
		\epBIC(\bttau)=eAS(\bttau)-c\tilde{m}\log n,
\end{equation*}
where the constant $c$ is user-defined. Here, we study its choice empirically. We use \textsc{g.WBS} here and all the other parameters are set at their default values.
Each simulation settings is repeated 1000 times, with dimensions $d=20, 50, 100, 500$ and 1000. The number of truly and falsely detected change-points are plotted in Figure \ref{fig:optiC}. A true change-point $\tau_j$ is deemed to be detected if an estimated change-point exists within 2 observations of it. The number of falsely detected change-points is defined as the number of estimated change-points minus the number of detected true change-points. We use $\llbracket a,b \rrbracket$ to represent the set of integers between $a$ and $b$ and $I$ to represent the identity matrix.
The following models are used to generate the data. 
\begin{itemize}
	\item Model 1: \begin{equation*}
		\bfy_i \sim
		\begin{cases}
			\mathcal{N}_{d}(\mathbf{0}, I) & \text{if } i \in \llbracket 1, 20\rrbracket \cup \llbracket 41, 60\rrbracket  \cup \llbracket 81, 100\rrbracket,\\
			\mathcal{N}_{d}(\frac{5}{4\log{(d)}}\boldsymbol{1}, I) & \text{if } i \in \llbracket 21, 40\rrbracket \cup \llbracket 61, 80\rrbracket  \cup \llbracket 101, 120\rrbracket.
		\end{cases}
	\end{equation*}
		\item Model 2: \begin{equation*}
		\bfy_i \sim
		\begin{cases}
			\mathcal{N}_{d}(\mathbf{0}, I) & \text{if } i \in \llbracket 1, 30\rrbracket \cup \llbracket 61, 90\rrbracket  \cup \llbracket 121, 150\rrbracket,\\
			\mathcal{N}_{d}(\mathbf{0}, (1+\frac{2}{\sqrt{d}})I) & \text{if } i \in \llbracket 31, 60\rrbracket \cup \llbracket 91, 120\rrbracket  \cup \llbracket 151, 180\rrbracket.
		\end{cases}
	\end{equation*}
		\item Model 3: \begin{equation*}
		\bfy_i \sim \begin{cases}
			 \text{Cauchy}_d(\mathbf{0},I) & \text{if } i \in \llbracket 1, 50 \rrbracket \cup \llbracket 101, 150 \rrbracket \cup \llbracket 201, 250 \rrbracket,\\
			 \text{Cauchy}_d(\frac{7}{4\log{(d)}}\boldsymbol{1},I) & \text{if } i \in \llbracket 51, 100 \rrbracket \cup \llbracket 151, 200 \rrbracket \cup \llbracket 251, 300 \rrbracket.
		\end{cases}
	\end{equation*}
		\item Model 4: \begin{equation*}
		\bfy_i \sim
		\begin{cases}
			\mathcal{N}_{d}(\mathbf{0}, \Sigma) & \text{if } i \in \llbracket 1, 40\rrbracket \cup \llbracket 81, 120\rrbracket  \cup \llbracket 161, 200\rrbracket,\\
			\mathcal{N}_{d}(\frac{1}{\log{(d)}}\boldsymbol{1}, I) & \text{if } i \in \llbracket 41, 80\rrbracket \cup \llbracket 121, 160\rrbracket  \cup \llbracket 201, 240\rrbracket,
		\end{cases}
	\end{equation*}
		where $\Sigma_{jk}=0.3^{\lvert j-k\rvert}$.

\end{itemize}

We can see from Figure \ref{fig:optiC} that the average true discoveries decreases faster after roughly $c=2$. For average false discoveries, it decreases slower after roughly $c=2$. Though the exact position of change varies a bit across different settings, we recommend $c=2$ since it reaches a good balance between power and false discovery rate.

\subsection{Result Visualization}
\label{hierarchicalstructure}
Given estimated change-points $\bhtau$, an essential question for researchers is whether these change-points are of scientific interest and what the relationship is between those subsegments. 
In our work, a visualization tool is provided to explore the hierarchical structure of the change-points.

In each step of Algorithm \ref{alg:gBE} \textsc{g.BE}, a suspicious change-point is removed, which is equivalent to merging two neighboring subsequences. The procedure progressively merges neighboring subsequences until all of them are in one big subsequence. So, the merging procedure could be viewed as a constrained hierarchical clustering process. 
In the beginning, each subsequence split by $\bhtau$ is in its own cluster. Then neighboring subsequences are merged as change-point is eliminated according to the criterion stated before. Given $\bttau^k, k=1, \dots, \hat{m}$ and its corresponding ep-BIC value, a change-point dendrogram is the ideal way to present the result of Algorithm \ref{alg:gBE}. 
The bottom of the change-point dendrogram is $\bhtau$, and the height in the dendrogram is negative ep-BIC. 

\begin{figure}[h!]
	\centering
	\includegraphics[width=0.65\linewidth]{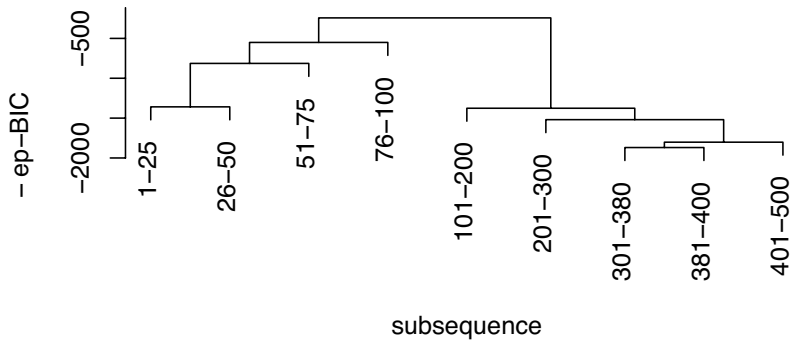}
	\caption{\footnotesize Change-point dendrogram constructed on a simulated data. The data contains 7 change-points, i.e., $\btau=\{ 25, 50, 75, 100, 200, 300,  400\}$. Among those detected change-points, 380 is a falsely detected one. The dendrogram shows that 380 is suspicious as adding 380 increases little in ep-BIC. }
	\label{fig:dendroeg}
\end{figure}

The tree structure of a change-point dendrogram depicts the relationship between estimated change-points and their relative importance (Figure \ref{fig:dendroeg}). If, for example, removing a given change-point results in minimal change in height, that change-point should be considered less significant or even doubtful. In contrary, a change-point that leads to a considerable ep-BIC lose is usually more locally important. For the change-points that are close to the root of the dendrogram, they are usually globally important. These change-points are removed at the end of backward elimination, which shows their importance in maintaining a high ep-BIC. In other words, these change-points cuts the inhomogeneous sequence into roughly homogeneous segments in a best effort with a fixed (small) number of change-points. 

One advantage of this hierarchical representation is that cutting the tree at a certain height will give a partitioning clustering at a corresponding level. This provides researchers with the freedom of choosing different scales of study. Often when dealing with complex data, it's often more important to grasp key changes than all of them. If so, one can cut the tree close to the top to get the top-level structure of the data. This is especially helpful when the data sequence is long and full of change-points.

\section{Performance Analysis}
\label{simulation}
In this section, we examine the performance of gMulti against two state-of-the-art nonparametric multivariate multiple change-point detection methods: E-Divisive \citep{matteson2014} and KCP \citep{arlot2019} implemented by the R package \texttt{ecp} \citep{Nicholas2015}.

Throughout the simulation, we set the number of intervals $L=100$, decay parameter $\gamma=\sqrt{0.5}$, significance level $\alpha=0.01$, and $\text{MinLen}=10$. 
Similarity graphs used are $\min (30, \lfloor \sqrt{b_l-a_l} \rfloor)$-MST in step 1 and $\min(5, \lfloor\sqrt{ \tilde{\tau}_{j+1}-\tilde{\tau}_{j-1}}\rfloor)$-MST in step 2. For the E-Divisive approach, we set the minimum cluster size to $\lfloor\min (\tau_{j+1}-\tau_j)/2\rfloor$ and all other parameters to default. For KCP, the maximum number of change-point is set to $2m$. 

\begin{remark}
	The default minimum cluster size for the E-Divisive is 30, which is greater than the minimum cluster size in our simulation settings. To make a fair comparison, we set it to $\lfloor\min (\tau_{j+1}-\tau_j)/2\rfloor$, though the results are barely affected. 
\end{remark}

The three methods are tested under five settings, with dimensions $d=20, 50, 100, 500$, and 1000 for Euclidean data settings, and the setting for network data detailed below. The number of truly and falsely detected change-points are reported in Table \ref{tab:comparisontd} and \ref{tab:comparisonfd}, respectively. When there is a mean change, only the first $d/5$ entries of the mean vector differ to make the change less significant. We define $\boldsymbol{\theta}$ as a sparse vector with the first $d/5$ entries equal to 1 and all others equal to 0. Location and scale parameters $(\delta, \sigma)$ are chosen for each value of $d$ so that most methods have moderate power.

The following models are used to generate the data. 
\begin{itemize}
	\item Model 5: \begin{equation*}
		\bfy_i \sim
		\begin{cases}
			\mathcal{N}_{d}(\mathbf{0}, \Sigma) & \text{if } i \in \llbracket 1, 50\rrbracket \cup \llbracket 101, 150\rrbracket  \cup \llbracket 201, 250\rrbracket,\\
			\mathcal{N}_{d}(\delta\boldsymbol{\theta}, \sigma\Sigma) & \text{if } i \in \llbracket 51, 100\rrbracket \cup \llbracket 151, 200\rrbracket  \cup \llbracket 251, 300\rrbracket,
		\end{cases}
	\end{equation*}
	where $\Sigma_{jk}=0.3^{\lvert j-k\rvert}$.
	\item Model 6: \begin{equation*}
		\bfy_i \sim \begin{cases}
			 \text{Cauchy}_d(\mathbf{0},I) & \text{if } i \in \llbracket 1, 40 \rrbracket \cup \llbracket 91, 145 \rrbracket \cup \llbracket 191, 255 \rrbracket,\\
			 \text{Cauchy}_d(\delta\boldsymbol{\theta},\Sigma) & \text{if } i \in \llbracket 41, 90 \rrbracket \cup \llbracket 146, 190 \rrbracket \cup \llbracket 256, 300 \rrbracket,
		\end{cases}
	\end{equation*}
	where $\Sigma_{jk}=0.3^{\lvert j-k\rvert}$.
	\item Model 7: \begin{equation*}
		\bfy_i \sim
		\begin{cases}
			\mathcal{N}_{d}(\mathbf{0}, I) & \text{if } i \in \llbracket 1, 55\rrbracket \cup \llbracket 91, 140\rrbracket  \cup \llbracket 196, 255\rrbracket,\\
			\mathcal{N}_{d}(\mathbf{0}, \sigma\Sigma) & \text{if } i \in \llbracket 56, 90\rrbracket \cup \llbracket 141, 195\rrbracket  \cup \llbracket 256, 300\rrbracket,
		\end{cases}
	\end{equation*}
	where $\Sigma_{jk}=0.3^{\lvert j-k\rvert}$.
	\item Model 8: \begin{equation*}
		\bfy_i \sim \begin{cases}	
		\text{Cauchy}_d(\mathbf{0},I) & \text{if } i\in \llbracket 1, 50 \rrbracket,\\
		\mathcal{N}_{d}(\frac{7}{\log(d)}\boldsymbol{\theta},\Sigma) & \text{if } i \in  \llbracket51, 65 \rrbracket,\\
		\text{Cauchy}_d(\mathbf{0}, 2I)  & \text{if } i \in \llbracket 66, 110 \rrbracket,\\
		\mathcal{N}_{d}(\frac{-5}{2\log(d)}\boldsymbol{\theta},I)& \text{if } i \in \llbracket 111, 160\rrbracket,\\
		\mathcal{N}_{d}(\frac{2}{\log(d)}\boldsymbol{\theta},\Sigma)& \text{if } i \in \llbracket 161, 185 \rrbracket,\\
		\text{Exp}_d(1)-1 & \text{if } i \in \llbracket 186, 260 \rrbracket,
 \end{cases}
	\end{equation*}
	where $\Sigma_{jk}=0.8^{\lvert j-k\rvert}$, and $\text{Exp}_d(1)$ represents $d$ dimensional independent exponential distribution with rate parameter 1.
	\item Model 9: A sequence of $n=240$ random networks are generated from the configuration model. To be specific, all nodes in a random graph have degree 2 if $i\in \llbracket 1, 30 \rrbracket \cup \llbracket 71, 115 \rrbracket \cup \llbracket 151, 205 \rrbracket $, otherwise the first 4 nodes have degree 4 and the others have degree 2. Let  $\bfy_i$ be the vectorized adjacency matrix of the $i$-th network. The dissimilarity is defined as $\lVert \bfy_i -\bfy_j \rVert_2$ for gMulti\footnote{We embedded the networks in the Euclidean distance so that all the methods in the comparison can be applied, while our method does not need such embedding.}.
\end{itemize}

\begin{table}[h!]
\footnotesize
 \caption{Number of detected true change-points and corresponding standard deviations (in parenthesis) with 1000 replications. The highest power are made bold.}
 \label{tab:comparisontd} 
\begin{center}
 \begin{tabular}{llccccc}
  \hline
  \hline
\multirow{6}{*}{Model 5}		& $d$ 				& 20 &50 & 100 & 500 & 1000\\
											 	& $(\delta, \sigma)$ & $(0.6, 1.85)$ &$(0.45, 1.75)$& $(0.37, 1.55)$ & $(0.1, 1.4)$ & $(0.05, 1.35)$\\
\cline{2-7}
												& gMulti (\textsc{g.WBS}) 	& 3.70 (0.99)	& \textbf{4.29} (0.79)	& \textbf{4.39} (0.73)	& \textbf{4.91} (0.34)	& \textbf{4.97} (0.19)\\
											 	& gMulti (\textsc{g.SBS}) 		& \textbf{3.75} (0.96)	& 4.14	(0.86)	& 4.21 (0.80)	& 4.74 (0.50)	& 4.91 (0.28)\\
												& E-Divisive							& 2.81 (1.35)	& 3.83 (1.01)	& 3.88 (1.02)	& 3.66 (1.17)	& 4.00 (1.00)\\
												& KCP										& 3.01 (1.83)	& 4.03 (1.46)	& 3.99 (1.43)	& 3.80 (1.75)	& 3.90 (1.71)\\
\hline
\multirow{6}{*}{Model 6}	& $d$ 				& 20 &50 & 100 & 500 & 1000\\
											& 	$\delta$ 		& 1.1 &0.85 & 0.76  & 0.64 & 0.6\\
\cline{2-7}
											& gMulti (\textsc{g.WBS}) 		& \textbf{3.35} (1.12)	& \textbf{3.34} (1.14)	& \textbf{3.50} (1.14)	& 3.41 (1.13)	& 3.21 (1.18)\\
											 & gMulti (\textsc{g.SBS}) 		& 3.31 (1.12)	& 3.27 (1.16)	& 3.39 (1.15)	& \textbf{3.53} (1.10)	& \textbf{3.44} (1.17)\\
											& E-Divisive							& 0.01 (0.09)	& 0.01 (0.08)	& 0.01 (0.09)	& 0.01 (0.09)	& 0.01 (0.08)\\
											& KCP										& 0.58 (0.85)	& 0.46 (0.75)	& 0.42 (0.70)	& 0.38 (0.71)	& 0.47 (0.70)\\
\hline
\multirow{6}{*}{Model 7}	& $d$ 				& 20 &50 & 100 & 500 & 1000\\
											& 	$\sigma$ 		& 1.9 &1.65 & 1.45  & 1.2 & 1.15\\
\cline{2-7}
											& gMulti (\textsc{g.WBS}) 		& \textbf{3.58} (1.00)	& \textbf{3.99} (0.93)	& \textbf{4.04} (0.92)	& \textbf{4.22} (0.84)	& \textbf{4.27} (0.82)\\
											 & gMulti (\textsc{g.SBS}) 		& 3.56 (1.04)	& 3.96 (0.93)	& 3.90 (0.94)	& 4.09 (0.87)	& 4.14 (0.87)\\
											& E-Divisive							& 0.64 (0.93)	& 0.81 (1.07)	& 0.27 (0.54)	& 0.10 (0.35)	& 0.04 (0.20)\\
											& KCP										& 3.03 (1.59)	& 2.65 (1.85)	& 1.29 (1.57)	& 1.34 (0.98)	& 1.00 (0.81)\\
\hline
\multirow{5}{*}{Model 8}	& $d$ 				& 20 &50 & 100 & 500 & 1000\\
\cline{2-7}
											& gMulti (\textsc{g.WBS})			& 4.17 (0.65)	& \textbf{4.09} (0.63)	& \textbf{4.03} (0.66)	& \textbf{3.95} (0.97)	& \textbf{3.69} (1.07)\\
											& gMulti (\textsc{g.SBS}) 		& \textbf{4.20} (0.64)	& \textbf{4.09} (0.65)	& \textbf{4.03} (0.67)	& 3.84 (1.02)	& 3.47 (1.09)\\
											& E-Divisive							& 1.33 (1.11)	& 0.88 (0.79)	& 0.69 (0.62)	& 0.66 (0.52)	& 0.64 (0.52)\\
											& KCP										& 1.36 (1.75)	& 0.96 (1.48)	& 1.09 (1.54)	& 0.89 (1.37)	& 0.86 (1.37)\\
\hline
\multirow{5}{*}{Model 9}	& number of nodes & 20 &30 & 50 & 75 & 100\\

\cline{2-7}
											& gMulti (\textsc{g.WBS})			& 4.84 (0.40)	& 4.90 (0.32)	& 4.93 (0.26)	& 4.93 (0.25)	& 4.95 (0.22)\\
											& gMulti (\textsc{g.SBS})		& \textbf{4.87} (0.35)	& \textbf{4.92} (0.28)	& \textbf{4.95} (0.22)	& \textbf{4.96} (0.18)	& \textbf{4.96} (0.21)\\
											& E-Divisive							& 4.78 (0.45)	& 3.73 (1.39)	& 0.87 (1.25)	& 0.23 (0.60)	& 0.10 (0.35)\\
											& KCP										& 3.27 (2.21)	& 4.14 (0.88)	& 2.78 (1.21)	& 1.92 (1.11)	& 1.56 (0.95)\\
\hline
 \end{tabular}
 \end{center}
\end{table}

\begin{table}[h!]
\footnotesize
 \caption{Number of falsely detected change-points and corresponding standard deviations (in parenthesis) with 1000 replications. The lowest false detection are made bold.}
 \label{tab:comparisonfd} 
\begin{center}
 \begin{tabular}{llccccc}
  \hline
  \hline
\multirow{6}{*}{Model 5}	& $d$ 				& 20 &50 & 100 & 500 & 1000\\
											& $(\delta, \sigma)$ & $(0.6, 1.85)$ &$(0.45, 1.75)$& $(0.37, 1.55)$ & $(0.1, 1.4)$ & $(0.05, 1.35)$\\
\cline{2-7}

 											& gMulti (\textsc{g.WBS}) 	& 1.80 (1.25)	& 1.46 (1.14)	& 1.41 (1.08)	& 0.99 (0.98)	& 0.90 (0.97)\\
 											& gMulti (\textsc{g.SBS}) 		& \textbf{1.39} (1.04)	& \textbf{1.07} (0.97)	& \textbf{1.09} (0.94)	& \textbf{0.72} (0.83)	& \textbf{0.43} (0.61)\\
											& E-Divisive							& 1.52 (1.10)	& 1.14 (0.97)	& 1.11 (0.98)	& 1.15 (1.01)	& 0.96 (0.93)\\  
											& KCP										& 2.74 (2.73)	& 3.82 (2.54)	& 5.00 (1.83)	& 4.62 (2.09)	& 4.63 (2.00)\\
\hline
\multirow{6}{*}{Model 6}	& $d$ 				& 20 &50 & 100 & 500 & 1000\\
											& $\delta$ 		& 1.1 &0.85 & 0.76  & 0.64 & 0.6\\
\cline{2-7}
											& gMulti	(\textsc{g.WBS})	& 1.75 (1.23)	& 1.67 (1.21)	& 1.52 (1.27)	& 2.06 (1.86)	& 2.46 (2.15)\\
											& gMulti (\textsc{g.SBS}) 		& 1.56 (1.11)	& 1.49 (1.05)	& 1.34 (1.07)	& 1.39 (1.16)	& 1.56 (1.35)\\
											& E-Divisive							& \textbf{0.09} (0.37)	& \textbf{0.07} (0.34)	& \textbf{0.08} (0.37)	& \textbf{0.07} (0.31)	& \textbf{0.06} (0.30)\\  
											& KCP										& 4.57 (4.44)	& 4.71 (4.54)	& 4.57 (4.57)	& 4.48 (4.59)	& 5.67 (4.47)\\
\hline
\multirow{6}{*}{Model 7}	& $d$ 				& 20 &50 & 100 & 500 & 1000\\
											& 	$\sigma$ 		& 1.9 &1.65 & 1.45  & 1.2 & 1.15\\
\cline{2-7}
											& gMulti (\textsc{g.WBS}) 		& 1.93 (1.23)	& 1.71 (1.18)	& 1.64 (1.20)	& 1.61 (1.19)	& 1.51 (1.13)\\
											 & gMulti (\textsc{g.SBS}) 		& 1.60 (1.07)	& 1.30 (1.03)	& 1.34 (1.04)	& 1.21 (0.98)	& 1.18 (0.98)\\
											& E-Divisive							&  \textbf{0.70} (0.96)	&  \textbf{0.82} (1.04)	&  \textbf{0.45} (0.78)	&  \textbf{0.23} (0.57)	&  \textbf{0.15} (0.46)\\
											& KCP										& 5.08 (2.67)	& 4.49 (2.97)	& 3.55 (3.76)	& 8.57 (1.00)	& 8.90 (0.84)\\
\hline
\multirow{5}{*}{Model 8}	& $d$ 				& 20 &50 & 100 & 500 & 1000\\
\cline{2-7}
											& gMulti	(\textsc{g.WBS})	& 0.21 (0.45)	& 0.30 (0.53)	& 0.41 (0.63)	& 1.45 (1.31)	& 2.03 (1.53)\\
											& gMulti (\textsc{g.SBS}) 		& \textbf{0.18} (0.40)	& \textbf{0.28} (0.53)	& \textbf{0.35} (0.57)	& 0.87 (0.99)	& 1.25 (1.16)\\
											& E-Divisive							& 0.27 (0.59)	& 0.39 (0.68)	& 0.41 (0.71)	& \textbf{0.37} (0.68)	& \textbf{0.38} (0.69)\\  
											& KCP										& 0.53 (1.58)	& 0.45 (1.51)	& 0.77 (1.98)	& 0.93 (2.23)	& 0.98 (2.29)\\
\hline
\multirow{5}{*}{Model 9}	& number of nodes & 20 &30 & 50 & 75 & 100\\

\cline{2-7}
											& gMulti	(\textsc{g.WBS})	& 1.09 (0.95)	& 1.11 (0.98)	& 0.94 (0.90)	& 0.79 (0.85)	& 0.69 (0.79)\\
											& gMulti (\textsc{g.SBS}) 		& 0.77 (0.83)	& \textbf{0.76} (0.79)	& 0.64 (0.73)	& 0.52 (0.67)	& 0.52 (0.66)\\
											& E-Divisive							& \textbf{0.25} (0.52)	& 0.85 (0.90)	& \textbf{0.55} (0.90)	& \textbf{0.27} (0.60)	& \textbf{0.16} (0.46)\\  
											& KCP										& 3.67 (2.48)	& 5.86 (0.88)	& 7.21 (1.21)	& 8.06 (1.11)	& 8.42 (0.95)	\\
\hline
 \end{tabular}
 \end{center}
\end{table}
 
From Table \ref{tab:comparisontd} and \ref{tab:comparisonfd}, we see that gMulti performs the best among these nonparametric methods under most simulation settings -- its power is higher than the other two methods, and its false discoveries are on the lower end. For cases where E-Divisive has a lower false discovery than gMulti, the power of E-Divisive is very low. Among the two implementation of gMulti, \textsc{g.SBS}-based gMulti has similar power and marginally better FDR compared to \textsc{g.WBS}-based gMulti. E-Divisive and KCP show comparable power under normal settings or low dimensions, but they can quickly fail under high dimension or some complex data structure. For KCP, it can have more falsely detected change-points than correctly detected ones, indicating its performance is greatly affected by dimension. These results show the robustness of gMulti compared to E-Divisive and KCP. The outstanding performance of the gMulti is partially due to the alleviation of curse-of-dimensionality by the generalized edge-count statistic.

\section{Real Data Analysis}
\label{realdata}
We illustrate the new framework on the Neuropixels data that record the activity of neurons in the brain of an awake mouse during spontaneous behaviors \citep{steinmetz2019}. The entire data can be found in \url{https://janelia.figshare.com/articles/dataset/Eight-probe_Neuropixels_recordings_during_spontaneous_behaviors/7739750/4}. The original data recorded the position and times of spikes using eight simultaneous Neuropixels probes across nine brain regions. For illustration, we consider the spike data for $d=176$ neurons in caudate-putamen during the first three minutes. The three minutes recording was discretized into $n=5400$ intervals of $1/30$ ms. Then $\bfy_{ij}$ represents the number of spikes recorded during time interval $i$ for neuron\footnote{Probes detected spikes in a small area of the brain that may cover more than one neuron. Here, we call this area neuron for simplicity.} $j$. Given the lack of an effective parametric model for such complex neural data, gMulti would be an appropriate choice for initial analysis. 

\begin{figure}[h]
	\centering
	\includegraphics[width=1\linewidth]{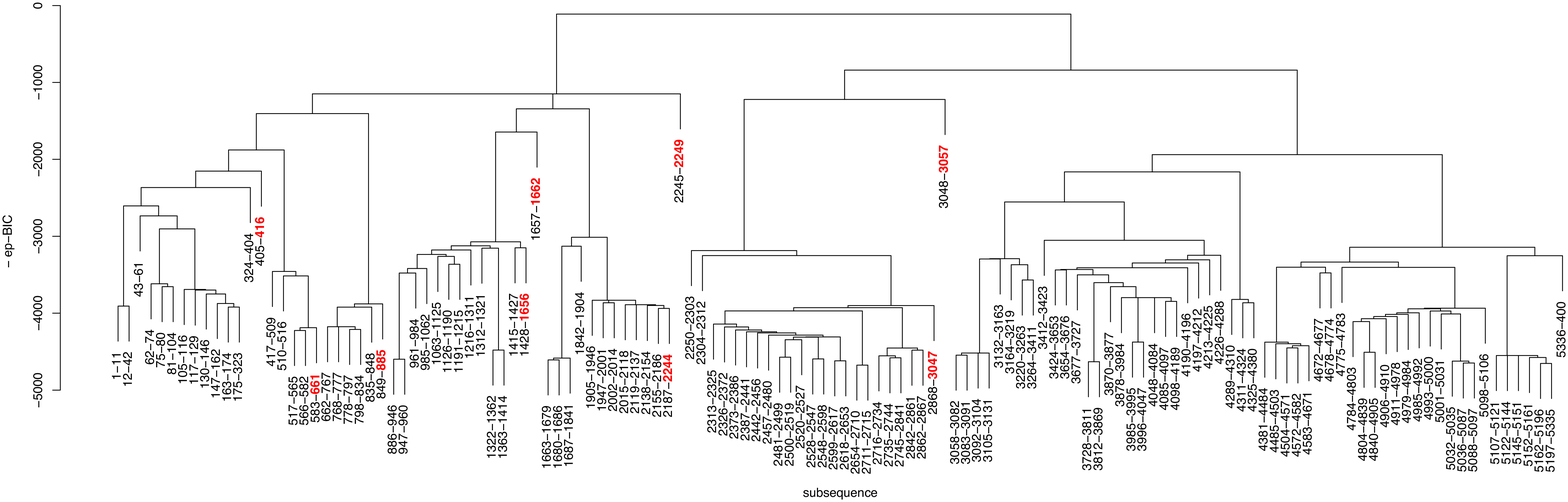}
	\caption{\footnotesize Change-point dendrogram of the Neuropixels recordings found by  \textsc{g.WBS}. There are 131 detected change-points. The maximum $\text{ep-BIC}=4674.29$. For better visualization, the hight of nodes are set to be at least the height of their children. The change-points in $\bttau^{9}$ are made bold.}
	\label{fig:dendroneuro}
\end{figure}

\begin{figure}[h]
	\centering
	\includegraphics[width=1\linewidth]{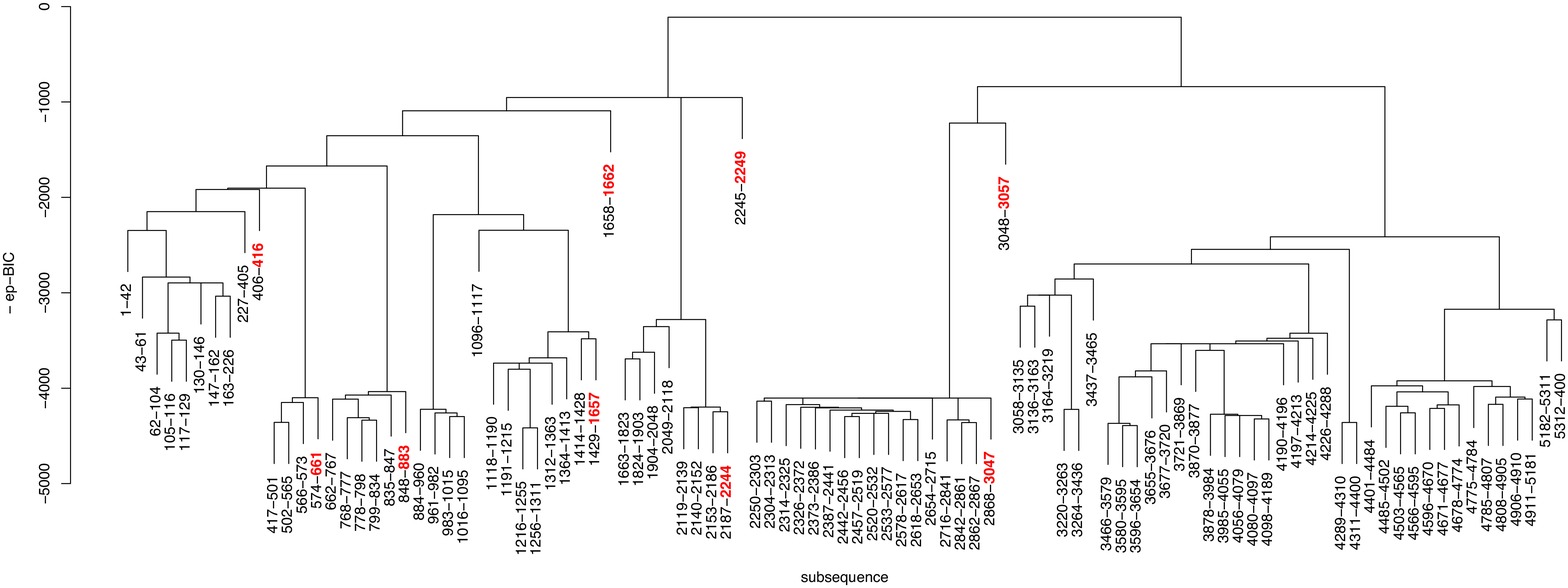}
	\caption{\footnotesize Change-point dendrogram of the Neuropixels recordings found by \textsc{g.SBS}. There are 98 detected change-points. The maximum $\text{ep-BIC}=4415.34$. For better visualization, the hight of nodes are set to be at least the height of their children. The change-points in $\bttau^{9}$ are made bold.}
	\label{fig:dendroneuroSBS}
\end{figure}

\begin{figure}[h]
\begin{center}
		\includegraphics[width=1\linewidth]{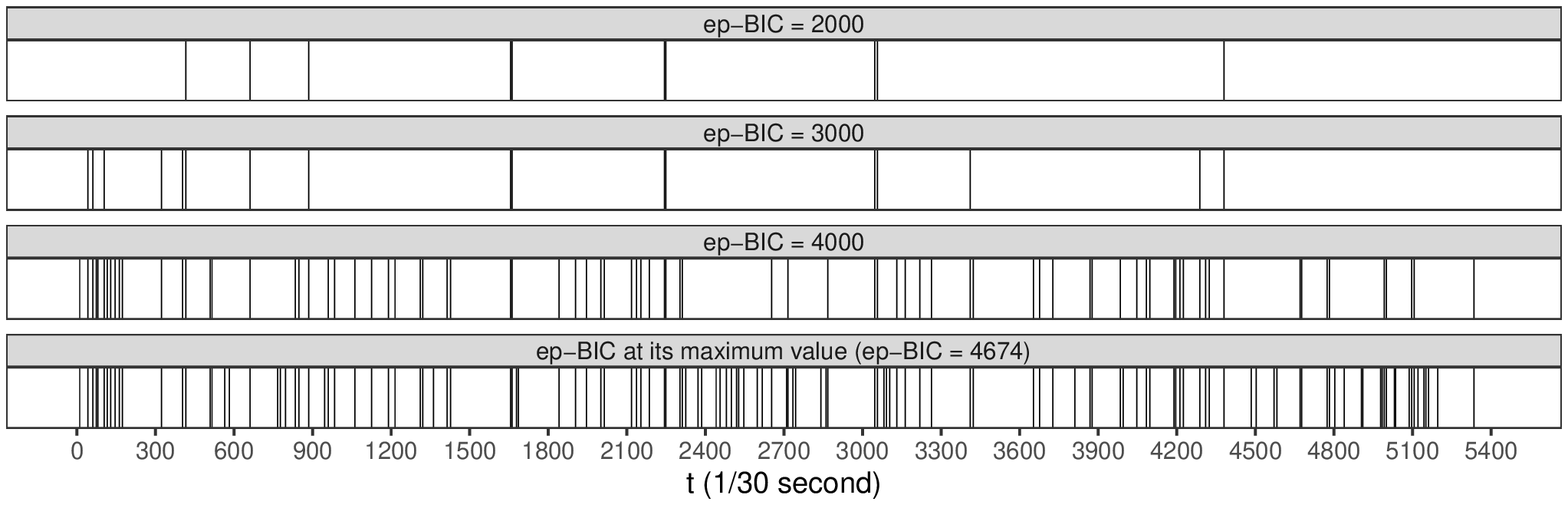}
\end{center}
	\caption{\footnotesize Position of detected change-points in \textsc{g.WBS}-based gMulti at four different $\text{ep-BIC}$ values. When $\text{ep-BIC}=2000$,  it corresponds to $\bttau^{10}=\{416, 661, 885, 1656, 1662, 2244, 2249, 3047, 3057, 4380\}$.} 
	\label{fig:neurobarcode}
\end{figure}
We use both \textsc{g.WBS}-based and \textsc{g.SBS}-based gMulti to analyze the sequence, with $\alpha=0.001$ to control local type I error and $L=200$ to ensure enough coverage for long sequences. Similarity graphs and other parameters are set to the default choices. 
The detected change-points are plotted in dendrograms in Figure \ref{fig:dendroneuro} and \ref{fig:dendroneuroSBS}. 
For \textsc{g.WBS}-based gMulti, in step 1, there are 258 candidate change-points detected, and 131 of them are kept after step 2, indicating frequent pattern changes in neural activities. For \textsc{g.SBS}-based gMulti, the two numbers are 181 and 98, respectively. Among those final 98 change-points in \textsc{g.SBS}-based gMulti, 73 of them are within 2 observations of change-points found by  \textsc{g.WBS}-based gMulti. The top level structure of the two methods are also very similar. The first 9 change-points in the  \textsc{g.WBS}-based and  \textsc{g.SBS}-based dendrograms are \{416, 661, 885, 1656, 1662, 2244, 2249, 3047, 3057\} and \{416, 661, 883, 1657, 1662, 2244, 2249, 3047, 3057\}, respectively (These change-points are in bold in Figure \ref{fig:dendroneuro} and \ref{fig:dendroneuroSBS}). 

\begin{figure}[p]
	\centering
	\includegraphics[width=0.9\linewidth]{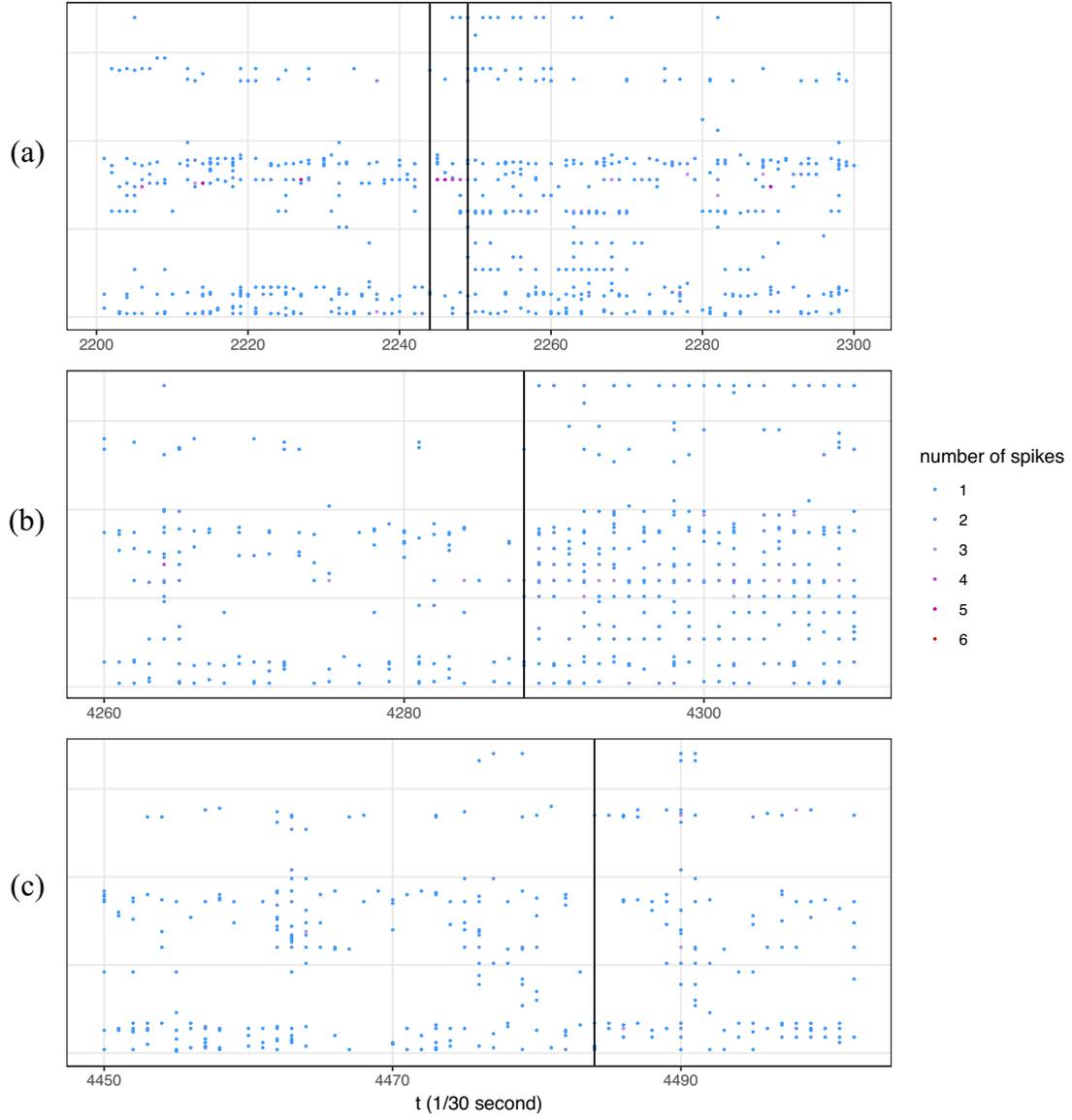}
	\caption{\footnotesize Neuropixel recordings and detected change-points. Vertical lines indicate positions of detected change-points. In (a), the 78th neuron is hyperactive for $i=2245,\dots, 2249$. In (b), most neurons are more active after $i=4288$. In (c), the covariance pattern changes after $i=4484$.}
	\label{fig:neurocpeg}
\end{figure}

\begin{figure}[h]
	\centering
	\includegraphics[width=0.8\linewidth]{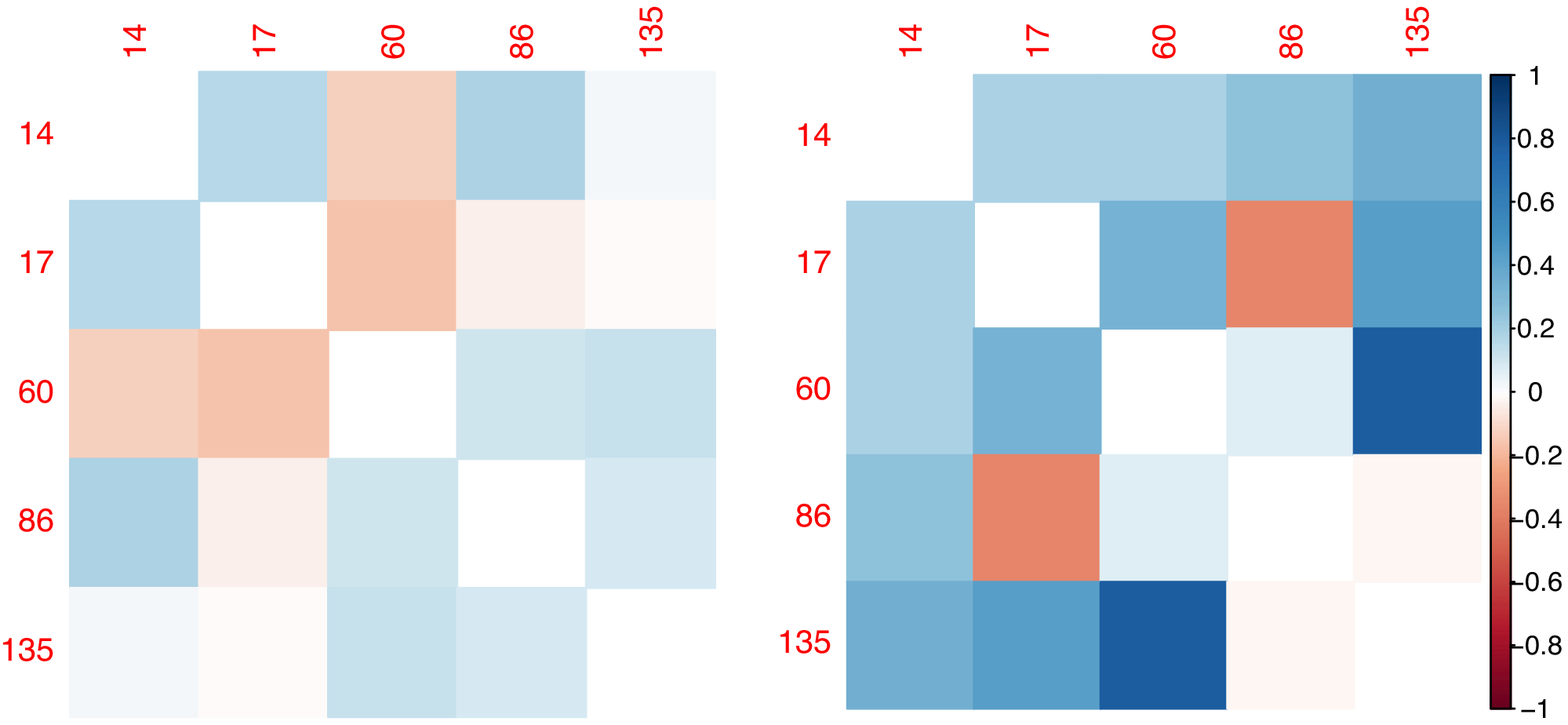}
	\caption{\footnotesize Correlation matrix of the five most active neurons before and after the change-point 4484. Indices of those neurons are also presented. The left penal shows the correlation matrix from $i=4450$ to 4484, and the right penal shows that from $i=4485$ to 4502.}
	\label{fig:corrplot}
\end{figure}

Figure \ref{fig:neurocpeg} plots some typical change-point patterns that might be of scientific interests. We call the first pattern \emph{single hyperactive neuron}, where a neuron suddenly becomes hyperactive for a short time interval. The hyperactivity can happen and disappear quickly and unexpectedly. An example is the 78th neuron for $i=2245$ to 2249 (Figure \ref{fig:neurocpeg} (a)). The second interesting pattern is \emph{overall intensity change}. This pattern is commonly seen in our data, like Figure \ref{fig:neurocpeg} (b), where most neurons are more (or less) active after the detected change-point at 4288. After an overall intensity change, the status can last for a long time until the next change-point. The third one is \emph{correlation pattern change}. This sometimes can happen together with overall intensity change. In Figure \ref{fig:neurocpeg} (c), the overall intensity barely changes after the change-point. If we use the overall number of spikes and perform the Mann-Whitney test, the $p$-value is 0.777. Nonetheless, if we calculate the correlation matrix of the five most active neurons before and after the change-point, we see that many neurons become more positively correlated after the change-point (Figure \ref{fig:corrplot}).

\section{Concluding Remarks}
\label{discussion}

In this paper, we focus on the generalized edge-count test. When there are some prior information, it could be that some other graph-based methods are more suitable. For example, the weighted edge-count test would be preferred if one is only interested in location alternatives. The arguments in this paper can be extended to the weighted edge-count test \citep{chen2018}. Let $Z_w^{[a,b]}(t)$ be the weighted edge-count scan statistic for $a \leq i \leq b$. Given the fact that $Z_w^{\left[a, b\right]}\left(t \right)^2 \stackrel{d}{\rightarrow} \chi_{1}^{2}$ under some regularity conditions, the corresponding extended pseudo-BIC may be defined as $\sum_{j=1}^{\tilde m}Z_w^{\left[\tilde\tau_{j-1}+1, \tilde\tau_{j+1} \right]}(\tilde \tau_j)^2-\tilde m \log n$. These goodness-of-fit statistics and detection algorithm may further be generalized to other nonparametric statistics. How to generalize them to other  statistics in a uniform framework is our next goal.

\bibliographystyle{agsm}
\bibliography{multicpbib}

\end{document}